\documentclass[letterpaper]{article} 
\usepackage{aaai2026}
\usepackage{times}  
\usepackage{helvet}  
\usepackage{courier}  
\usepackage[hyphens]{url}  
\usepackage{graphicx} 
\usepackage{natbib}  
\usepackage{caption} 
\usepackage{algorithm}
\usepackage{algorithmic}

\usepackage{newfloat}
\usepackage{listings}
\DeclareCaptionStyle{ruled}{labelfont=normalfont,labelsep=colon,strut=off} 
\floatstyle{ruled}
\newfloat{listing}{tb}{lst}{}
\floatname{listing}{Listing}
\title{Leader-driven or Leaderless:\\How Participation Structure Sustains Engagement and Shapes Narratives\\in Online Hate Communities}
\author{
    Rr. Nefriana\textsuperscript{\rm 1}\equalcontrib,
    Muheng Yan\textsuperscript{\rm 1}\equalcontrib,
    Rebecca Hwa\textsuperscript{\rm 2},
    Yu-Ru Lin\textsuperscript{\rm 1}\thanks{Corresponding author.}
}
\affiliations{
    \textsuperscript{\rm 1}University of Pittsburgh\\
    \textsuperscript{\rm 2}George Washington University\\

    rr.nefriana@pitt.edu, muheng.yan@pitt.edu, rebecca.hwa@email.gwu.edu, yurulin@pitt.edu 
}

\usepackage{bibentry}

\usepackage[most]{tcolorbox}
\definecolor{myred}{HTML}{D31010}
\newtcbox{\redcircled}{nobeforeafter, tcbox raise base,
  enhanced,
  colback=myred, colframe=myred, boxrule=0pt,
  coltext=white,
  width=2ex, height=2ex, 
  top=0pt, bottom=0pt, left=0pt, right=0pt,
  halign=center, valign=center,
  arc=1ex
}

\newtcolorbox{boxA}{
    colback = sub, 
    colframe = main, 
    boxrule = 1.5pt,
}

\newtcolorbox{boxB}{
    boxrule = 1.5pt,
}

\newtcolorbox{boxC}{
    colback = disclaimer, 
    boxrule = 1.5pt,
}

\usepackage{booktabs} 
\usepackage{multirow} 
\usepackage{placeins}

\usepackage{xcolor} 
\usepackage{soul} 

\definecolor{lightorange}{RGB}{255, 200, 120}
\definecolor{lightgreen}{RGB}{180, 255, 180}
\definecolor{lightblue}{RGB}{200, 230, 255}

\definecolor{softtorange}{RGB}{255, 230, 180}
\definecolor{disclaimer}{HTML}{FFE4EC} 

\sethlcolor{yellow}
\newcommand{\hlpink}[1]{\sethlcolor{pink}\hl{#1}\sethlcolor{yellow}} 

\newcommand{\hlgreen}[1]{\sethlcolor{lightgreen}\hl{#1}\sethlcolor{yellow}} 

\newcommand{\hllightblue}[1]{\sethlcolor{lightblue}\hl{#1}\sethlcolor{yellow}} 

\newcommand{\hlsoftorange}[1]{\sethlcolor{softtorange}\hl{#1}\sethlcolor{yellow}} 

\newcommand{\answerYes}[1]{\textcolor{blue}{#1}} 
\newcommand{\answerNo}[1]{\textcolor{teal}{#1}} 
\newcommand{\answerNA}[1]{\textcolor{gray}{#1}}

\begin{document}

\maketitle

\begin{abstract}
Extremist communities increasingly rely on social media to sustain and amplify divisive discourse. However, the relationship between their internal participation structures, audience engagement, and narrative expression remains underexplored. This study analyzes ten years of Facebook activity by hate groups related to the Israel–Palestine conflict, focusing on anti-Semitic and Islamophobic ideologies. Consistent with prior work, we find that higher participation centralization in online hate groups is associated with greater user engagement across hate ideologies, suggesting the role of key actors in sustaining group activity over time. Meanwhile, our narrative frame detection models—based on an eight-frame extremist taxonomy (e.g., dehumanization, violence justification)—reveal a clear contrast across hate ideologies: centralized Islamophobic groups employ more uniform messaging, while centralized anti-Semitic groups demonstrate greater framing diversity and topical breadth, potentially reflecting distinct historical trajectories and leader coordination patterns. Analysis of the inter-group network indicates that, although centralization and homophily are not clearly linked, ideological distinctions emerge: Islamophobic groups cluster tightly, whereas anti-Semitic groups remain more evenly connected. Overall, these findings clarify how participation structure may shape the dissemination pattern and resonance of extremist narratives online and provide a foundation for tailored strategies to disrupt or mitigate such discourse.


\end{abstract}


\section{Introduction}

\textcolor[HTML]{B22222}{Disclaimer: This paper contains offensive language for illustrative purposes; it does not reflect the authors’ views.} The rise of online hate and the spread of extremist rhetoric targeting Jewish and Muslim communities have become pressing global concerns \cite{chandra2021subverting, civila2020demonization, divon2022jewishtiktok, eckert2021hyper, ozalp2020antisemitism, vidgen2020detecting}. On platforms like Facebook, Muslim communities are dehumanized—described as ``filthy'' or the ``new Nazis"—while Jewish communities face Holocaust denial and victim-blaming narratives that delegitimize their historical suffering \cite{burke2020you}. These expressions of hate persist and often escalate, particularly during high-profile geopolitical events such as the 2023 escalation of Israel-Hamas conflict \cite{awan2016islamophobia, rose2021pandemic, adl2023online, nefriana2024shifting}. Victims report heightened anxiety and fear of real-world violence \cite{awan2017will}, and many engage in self-censorship to avoid harassment \cite{czymmek2022impact}.


While the harms of online hate are well-documented, research has only begun to examine certain aspects of how these communities function online. Prior studies point to the disproportionate influence of a few active users, the centrality of some individuals in networks, and the viral spread of memes across decentralized platforms \cite{vidgen2022islamophobes, goel2023hatemongers, zannettou2018origins, kasimov2025decentralized}. Yet, less is known about the structures that sustain these groups more effectively and how such structures relate to engagement dynamics or vary across ideological networks. Building on evidence from offline contexts showing that far-right groups led by charismatic figures or operating through leaderless structures are associated with mobilization toward violence \cite{chermak2013organizational, asal2020organizational}, we turn to online spaces to ask how different participation structures take shape in hate groups, what they imply for interaction within these communities, and how understanding them could inform more effective interventions.

We introduce the concept of \textit{participation structure} to capture the internal dynamics of online hate communities. Participation structure refers to the ways members produce and spread content, ranging from centralized—where a few influential figures dominate discussion—to decentralized, in which engagement is more evenly distributed across members. We then examine how these structures relate to engagement within anti-Semitic and Islamophobic Facebook groups. Engagement metrics, such as likes, comments, and shares, provide valuable insights into how effectively hate groups’ narratives resonate with their online audiences \cite{gallacher2021online}. Examining how variations in group dynamics shape patterns of engagement may illuminate why some hate groups persist longer and are harder to disrupt. 


Two contrasting theoretical frameworks that offer insight into how hate groups maintain cohesion inform our study. The \textbf{Leaderless Resistance} model describes decentralized, autonomous actors who operate independently to avoid detection and disruption \cite{beam1992leaderless, kaplan1997leaderless}. The \textbf{Club Goods} framework, drawing on Laurence Iannaccone and Eli Berman \cite{iannaccone1992sacrifice, berman2008religion}, emphasizes organized mechanisms that sustain loyalty and discipline within groups. Whether and how these offline patterns extend to online hate groups remains unclear.

To further explore how participation structures shape engagement patterns, we analyze the narratives circulating across groups with different structures and ideologies. Prior work on narrative framing shows that extremist groups strategically use rhetoric—such as claims of victimhood or ``us vs. them'' divisions—to reinforce identity and mobilization \cite{torregrosa2023survey}. We investigate which narratives dominate in each structure and ideology, and assess their variety. We also examine whether group structure relates to interaction patterns between groups: specifically, whether they tend to connect with ideologically similar groups (homophily), indicating more insular communication, or with ideologically different ones (heterophily), suggesting broader exchange.

Specifically, we derive four research questions:

\begin{itemize}
    \item \textbf{RQ1.} How do participation structures (centralized vs. decentralized) vary across hate groups with different ideological orientations (anti-Semitic vs. Islamophobic)?
    \item \textbf{RQ2.} How are participation structures linked to the intensity of engagement within online hate communities?
    \item \textbf{RQ3.} How are narratives framed and topics emphasized in centralized versus decentralized groups over time, and does having key actors in a group make narratives and topics more uniform or more varied?
    \item \textbf{RQ4.} How are participation structures related to inter-group connections—i.e., do groups engage more homophilically (with similar ideologies) or heterophilically (with different ideologies)?
\end{itemize}


By combining large-scale content analysis of 995,716 posts, Gini-based measures of participation centralization, narrative frame modeling, and network analysis, our findings indicate that higher participation centralization is associated with greater future engagement, suggesting that key actors might help sustain group activity over time. Participation structures also link to the development of hate narratives. Centralized Islamophobic groups exhibit more homogeneous messaging, while centralized anti-Semitic groups display broader framing and topical diversity—an aspect that also predicts higher future engagement levels. This contrast may stem from historical and political differences in each ideology’s development, as well as possible differences in leader coordination. Finally, although no significant relationship between centralization and homophily was observed, the groups displayed clear ideological differences, with Islamophobic groups clustering more tightly among themselves and anti-Semitic groups maintaining a relatively balanced distribution of connections both within and beyond their ideological network. These results offer practical entry points for intervention—for example, disrupting centralized leadership in coordinated Islamophobic groups and designing broader strategies that address both prominent figures and grassroots participants in more diffuse anti-Semitic communities. 

\section{Related Works}
\subsection{The Structures of Hate Groups}

In offline settings, hate group activities are often shaped by how they organize, exploit key events, and adapt to political contexts \cite{asal2020organizational}. Two competing theories explain which structures lead to greater success among hate groups. The leaderless resistance model suggests that groups with decentralized actors operating independently are more effective in avoiding detection and disruption \cite{beam1992leaderless, kaplan1997leaderless}. This approach has been used by white supremacist and jihadist groups to preserve flexibility \cite{stewart2024leaderless}. Supporting the superiority of this model, \citet{asal2020organizational} find that decentralized groups are more likely to commit violence. 

In contrast, extending the club goods framework \cite{iannaccone1992sacrifice}, Eli Berman argues that controlled and organized mechanisms—such as the provision of social services and the enforcement of internal norms—help maintain loyalty and discipline within successful radical religious groups \cite{berman2008religion}. Other scholarly works align with this theory, showing that strong leader-driven mobilization of ideology and resources supports group growth and longevity \cite{freilich2009critical}, and that many terrorist groups dissolve when they lose key members to arrest or death \cite{jones2008terrorist}.

Existing studies on online hate reveal patterns in how it spreads across centralized and decentralized networks. For example, a small number of highly active users produce most Islamophobic content on Twitter \cite{vidgen2022islamophobes}, and hate speakers often occupy central network positions and initiate cascades \cite{goel2023hatemongers}, indicating reliance on key actors. In contrast, anti-Semitic memes spread virally on platforms like 4chan, Reddit, and Gab, and adaptive link dynamics across smaller platforms further reinforce decentralized hate networks \cite{kasimov2025decentralized, zheng2024adaptive, zannettou2018origins}.

While these studies reveal user roles in centralized versus decentralized networks, they do not examine the internal structures that sustain hate groups’ activity over time or how these structures are associated with their narrative patterns and connectivity between groups. Although prior studies suggest that both centralized online mobilization—supported by leadership and formal organization—and online activism on a decentralized platform—driven by user dynamics—can foster high levels of interaction \cite{kasimov2025decentralized, poole2021tactical}, it remains unclear which structure is more effective, particularly in the context of Islamophobic and anti-Semitic groups. Moreover, ideological content—such as Islamophobic or anti-Semitic material—is often studied in isolation rather than in relation to, or in comparison with, other forms of hate. This study contributes to the literature by analyzing how hate groups’ internal structures relate to engagement patterns and vary across ideological networks.


\subsection{Narrative Framing in Hate Groups}
Hate groups, from Islamist extremists to right-wing radicals, use compelling narratives to attract, persuade, and radicalize individuals \cite{torregrosa2023survey}.
Storytelling in propaganda fosters identity, belonging, and purpose among recruits \cite{frischlich2018power}. 
These narratives emotionally resonate with the audience by framing their messages in relatable terms \cite{ferguson2020staying}.
Groups like ISIS, for instance, frame attacks as defensive actions to protect their religious community, portraying violence as a justified Jihad or holy struggle \cite{ferguson2020staying}. 
This framing not only aids recruitment but also retains member loyalty by presenting their cause as noble.

Hateful narratives use positive framing to justify their actions, often drawing on religious or moral values, historical references, and the glorification of members \cite{williams2022does}.
Conversely, negative framing demonizes outgroups, promoting an binary-confronting mentality that enhances ingroup identity and dehumanizes outsiders \cite{gerard2025fear, van2023producing,lilleker2023demonising}.
Such language desensitizes individuals to hate and strengthens group cohesion through shared identity and cognitive bias \cite{bouko2022discourse}.

Although the existing studies emphasize the importance of narration, narrators' intent, and framing in understanding the ecology of hateful propaganda, there has not been comprehensive research that recognizes these related concepts holistically.
Through the use of specific language and perspective, framing highlights certain aspects of a narrative while downplaying others, guiding the audience's interpretation and emotional response \cite{linDynamicsTwitterUsers2020}. 
In this study, we contribute by introducing a novel topology and coding scheme that capture the narration intents of centralized and decentralized hate groups. These tools allow us to generate analytical results that deepen our understanding of how such ideologies are articulated.

\section{Methodology}

\begin{figure*}[t]
    \centering
    \includegraphics[width=0.8\textwidth]{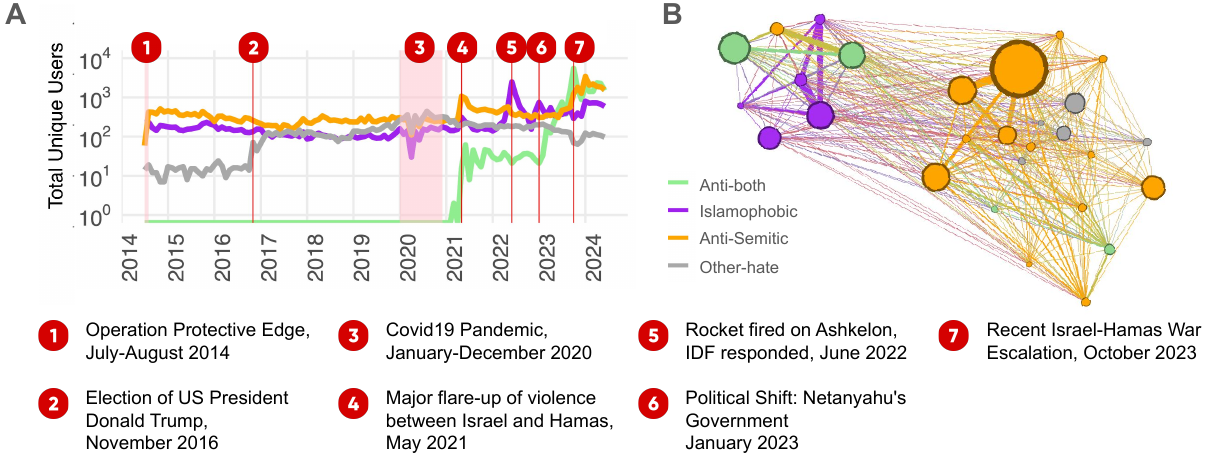} 
    \caption{A decade of data on online hate groups related to the Israel/Palestine conflict on Facebook: (A) total unique users over time across four hate group ideologies, (B) the overall network of online hate groups, with nodes representing groups, node size indicating post count, edges representing shared users, and edge weights corresponding to the number of shared users.}
    \label{fig:dataset}
\end{figure*}

\subsection{Dataset}
To identify the hate groups, we first used the CrowdTangle API \cite{crowdtangle2020data} to collect posts over a two-month period following the outbreak of the Israel-Hamas conflict on October 7, 2023. Keywords and phrases suggested by two political scientists who are experts in Middle Eastern politics included explicit hate speech and conflict-related terms such as ``death to Israel,'' ``Arabs are inferior,'' and ``From the Rivers to the Sea''. These terms are used by both Islamophobic and Anti-Semitic groups to express the desire to eliminate Jews/Muslims from the Jordan River to the Mediterranean Sea. 

This initial search identified 30 active groups, and three graduate students reviewed posts for hate speech targeting Jews, Muslims, or other groups/ideologies (e.g., White Supremacy). Posts flagged for hate speech were then reviewed by one author to finalize group labels: Anti-Semitic, Islamophobic, Anti-both (containing both Anti-Semitic and Islamophobic content), or Other-hate. Given the nuanced nature of hate group categorization and its relevance to the Israel-Palestine conflict, we broadly define Anti-Semitic to include hate directed at both Jewish communities and individuals associated with Israel. Similarly, Islamophobic sentiment encompasses prejudice against Islamic communities and those identified with Palestine. This broader categorization helps capture the complexity of these hate narratives while acknowledging limitations in the dataset.

After extending the data collection, our dataset spans from July 2014 to June 2024 (Fig. \ref{fig:dataset}). This longitudinal dataset includes 1,218,789 posts and 43,971 users from 28 of the 30 identified groups, as two were banned. For time-series analyses, we excluded the Anti-both category due to data sparsity (i.e., limited monthly observations) (Fig. \ref{fig:dataset}A). The final dataset consists of 995,716 posts and 43,971 unique users from the remaining 24 groups, yielding 1,820 monthly group observations (see Appendix ``Hate Group Distribution"). Of these, 55.63\% were media-only posts, and 13.19\% were in languages other than English. Textual analysis (including framing analysis and topic modeling) concentrated on the 310,532 English-language posts, which enabled us to examine narrative framing and content shifts across hate group structures and ideologies.

\subsubsection{Auxiliary Dataset}
To train a classifier for inferring narrative frames, we used an auxiliary dataset from prior work on hate speech \cite{gaikwad2021multi}. This dataset includes 3,864 White Supremacist texts from the StormFront forum. By combining with our Facebook dataset, we created a merged set of 940 samples, evenly split between the two sources. 


\subsection{Measurement}
This subsection describes methods for extracting measures, computed monthly for each group to track temporal change.

\subsubsection{Group Participation Structure/Centralization.} \label{sec:gini_user}
Unlike conventional groups, online groups often lack formal structures, which complicates the identification of leadership. Our study used participation centralization as a proxy, conceptualizing users who dominate content creation as leaders who may influence the spread of hate speech. To quantify centralization, we employed the Gini coefficient, a widely used measure of inequality suited to capturing disparities in fat-tailed posting activity (see Appendix ``Methodological Choices" for details). This allowed us to assess whether participation is concentrated among a few users (high Gini coefficient, suggesting centralization) or more evenly distributed (low Gini coefficient, indicating decentralization). 

\subsubsection{Content Homogeneity.} We also utilized the Gini coefficient to measure monthly framing and topic homogeneities based on the number of posts. However, instead of measuring posting rate inequality among users, we measured the inequality in the distribution of posts across various narrative framings (for framing homogeneity) and across different topics (for topic homogeneity). Here, a higher Gini value indicates greater content concentration, where the discourse is dominated by fewer frames or topics. In contrast, a lower Gini value indicates a more diverse discussion, with users distributing attention more evenly across frames and topics, as reflected in post counts. Comparing the Gini coefficients between centralized and decentralized groups can provide insights into whether (a) framings or topics of interest may be guided by group leaders (i.e., when framing/topic concentration is higher in centralized groups than in decentralized ones), or (b) framing/topic emphasis emerges collectively through group consensus (i.e., when framing/topic concentration is lower or more diverse in centralized groups than in decentralized ones). 

\subsubsection{Group Network Attributes.} To examine how hate groups interact with one another in an online space, we constructed monthly networks from our dataset, where each node represents a unique group, and an edge is established between two nodes based on shared users—those who posted in both groups. The weight of each edge reflects the number of users common to the two groups. Using these networks, we subsequently calculate the group degree and homophily index.


We measured hate groups’ degree—defined as the number of distinct groups they are affiliated with—by identifying all groups that share at least one user with the target group—that is, a user who posted in both groups. We focus on degree because previous research highlights the importance of intergroup connections: groups connected to more other groups are associated with higher fatalities \cite{asal2008nature} and a greater likelihood of attacks on vulnerable civilians \cite{asal2009softest}.


We measured homophily—defined as the tendency of a group to connect with other groups that share the same ideology (Islamophobic, Anti-Semitic, Anti-both, or Other-hate groups)—using a weighted Homophily Index ($HI$) adapted from \citeauthor{krackhardt1988informal}’s \citeyearpar{krackhardt1988informal} External–Internal index:
\newline
\[
    \text{$HI$} = \frac{W_\text{in} - W_\text{out}}{W_\text{in} + W_\text{out}}
\]
\newline
where $W_{in}$ is the internal weight, or the total connections to groups with the same ideology, and $W_{out}$  is the external weight, representing connections to groups with different ideologies. $HI = 1$ indicates full homophily (connections only to the same ideology) and $HI = -1$ indicates full heterophily (connections only to different ideologies). Our study focuses on weighted connections when calculating homophily since the strength of these connections helps us understand how tightly information is contained within groups and how it might spread to or influence other groups.

\subsection{Classification of Narrative Frames}

\subsubsection{Narrative Framing Taxonomy.}We developed the narrative framing taxonomy using Grounded Theory procedures \cite{glaser2017discovery}. We started with a partial taxonomy of narrative framing in extremist groups, proposed by \citeauthor{vandenberg2021legitimating} (\citeyear{vandenberg2021legitimating}), which includes five main frames: defensive, moralistic, legalistic, imperialistic, and apocalyptic. These frames legitimize political violence, particularly in jihadist organizations such as Al Qaeda and ISIS \cite{vandenberg2021legitimating}. However, this taxonomy omits other important frames, including: (1) in-group versus out-group binary separation \cite{bennett2011war, brindle2016language}, (2) demonization and dehumanization of others \cite{brindle2016language, ebner2024assessing}, and (3) romanticizing charismatic leaders or ideological pioneers \cite{mansouri2018introduction, meiering2020connecting}. We integrated these additional frames into the original five and, through axial and selective coding, identified eight key narrative frames based on theoretical and case study analyses.

\begin{itemize}
    \item \textbf{\textit{Us Vs. Them:}} establishes a binary distinction between the in-group (us) and out-group (them), creating a sense of polarization and conflict.
    \item \textbf{\textit{Heroic:}} positions the in-group as heroes fighting against out-groups, glorifying those who sacrifice for the cause and portraying them as symbols of commitment.
    \item \textbf{\textit{Dehumanization and Demonization:}} portrays the out-group as less than human, evil, or inherently corrupt.
    \item \textbf{\textit{Victimization:}} describes the in-group as victims of oppression, persecution, or injustice.
    \item \textbf{\textit{Justification of Violence:}} presents the hate group's beliefs as absolute moral truth, rejecting any form of compromise or dialogue.
    \item \textbf{\textit{Legitimacy:}} lends the narratives a semblance of credibility and authenticity, sometimes drawing on real or imagined past events to justify current actions and goals.
    \item \textbf{\textit{Imminent War/Crisis:}} describes situations as urgent, requiring immediate action, often portraying impending doom or war that can only be avoided through the group's proposed actions.
    \item \textbf{\textit{Religious:}} uses religious texts or beliefs to justify or promote extremist views or actions.
\end{itemize}

These eight narrative frames cover all cases in our Facebook and StormFront samples. Following the Grounded Theory procedures, we created a detailed narrative framing annotation codebook, including definitions, inclusion criteria, and examples from the dataset (see Appendix ``Extremist Narrative Framing Classification: Codebook").

\subsubsection{Establishing Ground Truth for Narrative Frames}

\paragraph{Annotation.} Three native English speakers were trained to apply a narrative framing codebook before annotating the texts. 
Each annotator independently evaluated the same set of texts and indicated all narrative frames that applied. They coded approximately 100 samples per week and met weekly to refine codebook use and resolve ambiguities. Pairwise Cohen’s Kappa scores ranged from 0.55 to 0.85, indicating substantial to high reliability. Final labels were assigned by majority votes. 

\paragraph{Sample Augmentation.} Preliminary classification experiments with the annotated dataset produced accuracy scores between 65--75\%, indicating challenges in training classifiers on sparse data. To increase coverage for underrepresented frames, we used Llama-2 for data augmentation because it offered controllable prompt-based generation, stable output quality, and reproducibility. While newer models have emerged, Llama-2 provided the transparency and control necessary to produce high-quality synthetic data for this research \cite{touvron2023llama}. Augmentation focused on four underrepresented narrative frames: \textit{Justification of Violence, Imminent War/Crisis, Legitimacy,} and \textit{Heroic}. 

Prompts were constructed using three human-annotated examples, the corresponding frame definition from the codebook, and an assigned ideological role (\textit{White Supremacist, Muslim-hater,} or \textit{Jew-hater}). The exact prompt used was:

\begin{quote}
\textit{``Here are examples of an extremism narrative: [Post 1], [Post 2], [Post 3]. The posts are under a [NF] framing, which is defined as: [Definition 1], [Definition 2]. Assume you are a [Ideology Role], can you write a different sample with the same [NF] framing?''}
\end{quote}

\noindent where \textit{[NF]} is the narrative frame, \textit{[Definition]} is the frame definition from the codebook, and \textit{[Ideology Role]} specifies the viewpoint from which the narrative should be written.

To evaluate the quality of the generated samples, annotators completed trials in which each trial presented four texts: two generated by Llama-2 intended to match the target frame, and two human-written samples from other frames. Annotators selected the two texts that best reflected the target frame definition. Across 640 trials, Precision@2 scores were 0.82, 0.86, and 0.72, indicating that most generated samples aligned with the intended frames. Samples that did not meet these criteria were discarded, and the validated texts were incorporated into the training corpus (see Table \ref{tab:narrative_samples} for the distribution of sample before and after augmentation).

\paragraph{Machine Classification of Narrative Frames.} After creating the labeled dataset, we developed a classification pipeline combining zero-shot learning with Llama-2 \cite{touvron2023llama} and supervised fine-tuning of a transformer model. As a baseline, Llama-2 was prompted with a text sample and the definitional criteria of a narrative frame. While it achieved strong recall, it often overestimated frame presence, resulting in lower precision. To improve performance, we fine-tuned a RoBERTa-large model \cite{Liu2019RoBERTa} that had previously been adapted to online hate speech corpora. The fine-tuned model achieved F1 scores ranging from 0.85 to 0.94, with an average improvement of 37.29\% over the zero-shot baseline (see Appendix Table A5 for results and Appendix ``Extremist Narrative Framing Classification" for more details).


\subsection{Identification of Discourse Topics} 

We used BERTopic \cite{grootendorst2022bertopic} to extract topics from the post.
After pruning the tree at a depth of three, we identified five major topic clusters, and manually assigned names based on the top keywords, as: \textit{Social Media Activism} (activism-related terms and hashtags), \textit{Military} (military actions and casualties), \textit{Ethnicity Hate/War Crime Charges} (accusations of ethnic-based hate or war crimes), \textit{Political/Protest} (discussions and protests, mainly from U.S. perspectives), and \textit{Religious} (religious references and arguments).

\subsection{Analysis Methods}
\paragraph{Answering RQ1.} Bootstrapped Mann–Whitney \textit{U} tests with Bonferroni correction were used to assess differences in Gini coefficient of user participation distributions across ideologies. See Appendix ``Methodological Choices" and ``Additional Analysis Notes" for justifications and details. 

\paragraph{Answering RQ2.}We used Negative Binomial regression models to examine the month-to-month relationship between groups’ participation structure and engagement across hate ideologies (RQ2), along with factors such as content homogeneity, network characteristics, and controls (including adjustments for serial correlation). Negative Binomial was chosen to account for the over-dispersed count outcome. See Appendix ``Methodological Choices" and Appendix ``Additional Analysis Notes: Regression Analysis" for detailed justification and specification.

\paragraph{Answering RQ3 and RQ4.} To investigate the relationship between user participation structure and other factors (i.e., framing homogeneity, topic homogeneity, and homophily, we dichotomized the monthly Gini coefficients of user participation across all online hate groups over a ten-year period. Groups were classified as decentralized ($\text{Gini ratio} \le \text{median}$) or centralized ($\text{Gini ratio} > \text{median}$)
(for the justification and sensitivity analysis, see Appendix ``Methodological Choices"). We then used bootstrapped Mann–Whitney \textit{U} tests with Bonferroni correction to assess differences in framing and topic homogeneity (RQ3) and homophily levels (RQ4) across hate group ideologies and between centralized and decentralized groups, reporting median and interquartile range statistics (see Appendix ``Additional Analysis Notes" for more details). 

\section{Results}


\subsection{Participation Structure Across Ideologies (RQ1)}

\paragraph{Summary of Findings:} \textit{ Regardless of ideological focus, a small subset of highly active members consistently dominates content production in hate groups.}

\begin{figure}[h]
    \centering
    \includegraphics[width=0.79\columnwidth]{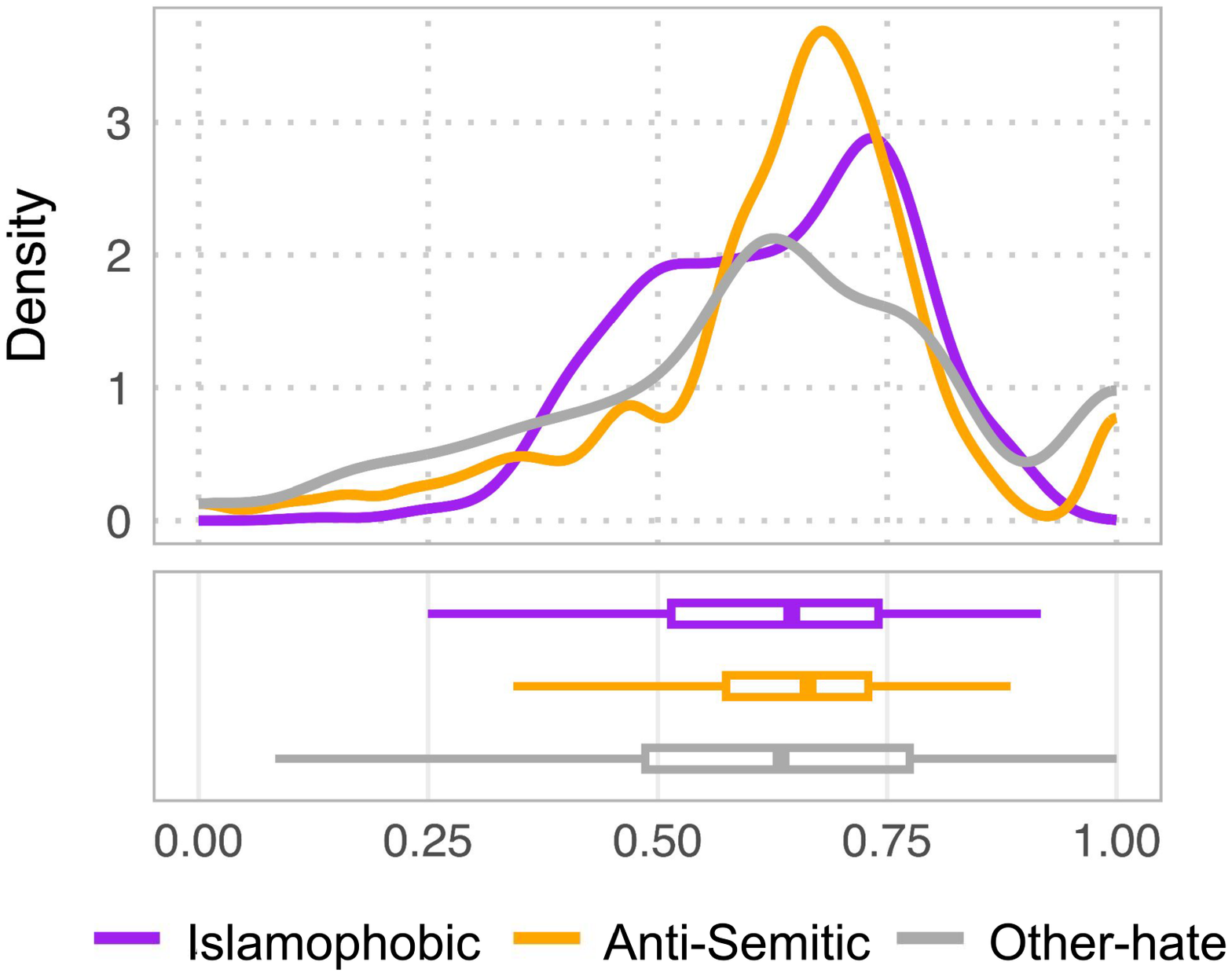} 
    \caption{Distributions of Gini coefficients measuring inequality of user participation across hate group ideologies. The plots reveal similar participation structures across the three ideologies. The left-skewed distributions indicate that a small proportion of users produce most group content.}
    \label{fig:gini_users}
\end{figure}

Fig. \ref{fig:gini_users} shows that the Gini Coefficient of user participation distribution is similarly left-skewed across all hate ideologies, indicating a prevalent, highly centralized pattern with content creation concentrated among few participants (Islamophobic: $median = 0.65$, $IQR = 0.23$; Anti-Semitic: $median = 0.66$, $IQR = 0.16$; Other-hate: $median = 0.63$, $IQR = 0.28$). This finding is consistent with \citeauthor{vidgen2022islamophobes} (\citeyear{vidgen2022islamophobes}), who reported that a small number of highly active users produced the majority of Islamophobic content on Twitter. Bootstrapped Mann--Whitney \textit{U} tests with the significance level adjusted for multiple comparisons indicate that differences in centralization among the ideologies are not statistically significant (see Appendix ``Additional Analysis Notes"). 
Despite ideological differences, these results demonstrate consistent structural similarities in participation dynamics across hate group ideologies.

\subsection{Participation Structure and Engagement Level (RQ2)}


\paragraph{Summary of Findings:}\textit{Across hate group ideologies, participation centralization is strongly associated with increased engagement.}

\begin{figure}[h]
    \centering
    \includegraphics[width=\columnwidth]{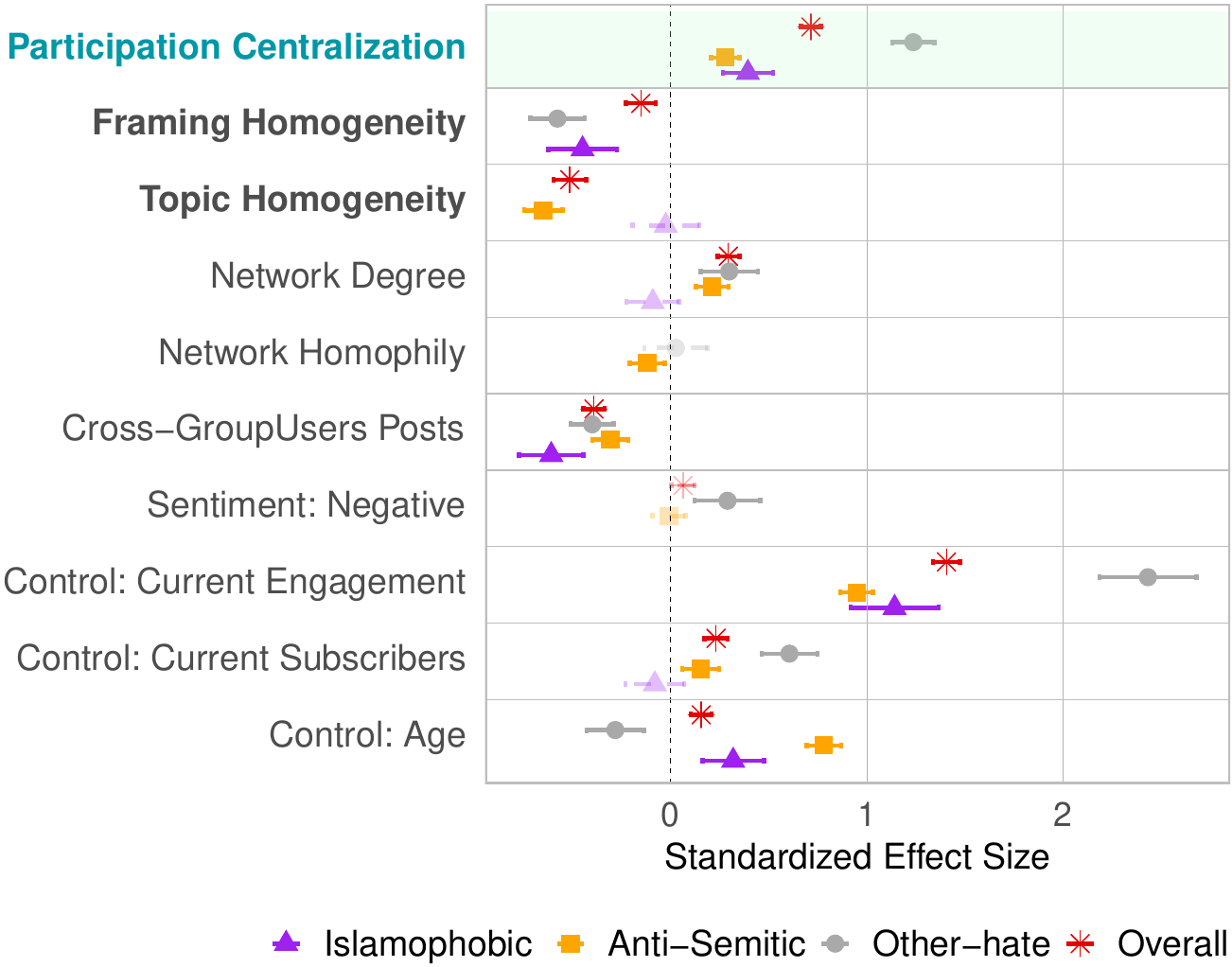} 
    \caption{Regression results predicting next-month engagement by group ideology. 
    Solid (as opposed to faded) markers show coefficients that are both 
    statistically significant ($p < 0.05$ and $95\%$ $CI$ does not include 0) and 
    practically significant ($|\beta| \geq 0.1$). Predictors were selected independently for each 
    ideological category based on multicollinearity ($VIF$) and model fit 
    (\textit{AIC}), resulting in ideology-specific models where not all predictors 
    are present across all groups. This figure shows that participation centralization is a significant predictor of engagement across ideologies. 
    Full regression results are provided in 
    Appendix.}
    \label{fig:regression_analysis}
\end{figure}

Fig. \ref{fig:regression_analysis} shows our regression model results for predicting next month's engagement level in online hate groups: Islamophobic ($R\textsuperscript{2} = 0.804$), Anti-Semitic ($R\textsuperscript{2} = 0.927$), Other-hate groups ($R\textsuperscript{2} = 0.973$), and overall ($R\textsuperscript{2} = 0.912$). In this specific analysis, the engagement of interest refers to the number of interactions such as likes and shares with content within the groups. The analysis shows that, both within each hate group ideology and in the aggregate, higher participation centralization is significantly associated with increased engagement (Islamophobic: $\beta_{centralization} = 0.39$, $p < 0.001$, $CI = [0.27, 0.52]$; Anti-Semitic: $\beta_{centralization} = 0.28$, $p < 0.001$, $CI = [0.21, 0.35]$; Other-hate: $\beta_{centralization} = 1.24$, $p < 0.001$, $CI = [1.13, 1.34]$; Overall: $\beta_{centralization} = 0.72$, $p < 0.001$, $CI = [0.67, 0.77]$).

Besides this main finding, we also found that, across all group ideologies, content homogeneity is significantly associated with lower engagement (Overall: $\beta_{framing Homogeneity} = -0.15$, $p < 0.001$, $CI = [-0.23, -0.08]$; $\beta_{topic Homogeneity} = -0.51$, $p < 0.001$, $CI = [-0.60, -0.43]$; Fig. \ref{fig:regression_analysis}). Yet, the relationship between content homogeneity and future engagement is not the same across ideologies. In Islamophobic and Other-hate groups, framing homogeneity shows a significant connection to lower engagement (Islamophobic:  $\beta_{framing Homogeneity} = -0.45$, $p < 0.001$, $CI = [-0.62, -0.28]$; Other-hate:  $\beta_{framing Homogeneity} = -0.58$, $p < 0.001$, $CI = [-0.72, -0.44]$). In Anti-Semitic groups, topic homogeneity appears to be more relevant in predicting future’s engagement ($\beta_{topic Homogeneity} = -0.65$, $p < 0.001$, $CI = [-0.75, -0.55]$).

\subsection{Participation Structures and The Narratives Within Hate Groups (RQ3)}

\paragraph{Summary of Findings:} \textit{The centralization of participation among a few key actors corresponds differently with narrative diversity across ideological categories. Centralized Islamophobic groups have narrower, more consistent narrative frames and topics, whereas decentralized Islamophobic groups show greater diversity. In contrast, centralized Anti-Semitic groups employ more varied narrative frames and topics than their decentralized counterparts.
}

\begin{figure*}[ht]
    \centering
    \includegraphics[width=1.0\textwidth]{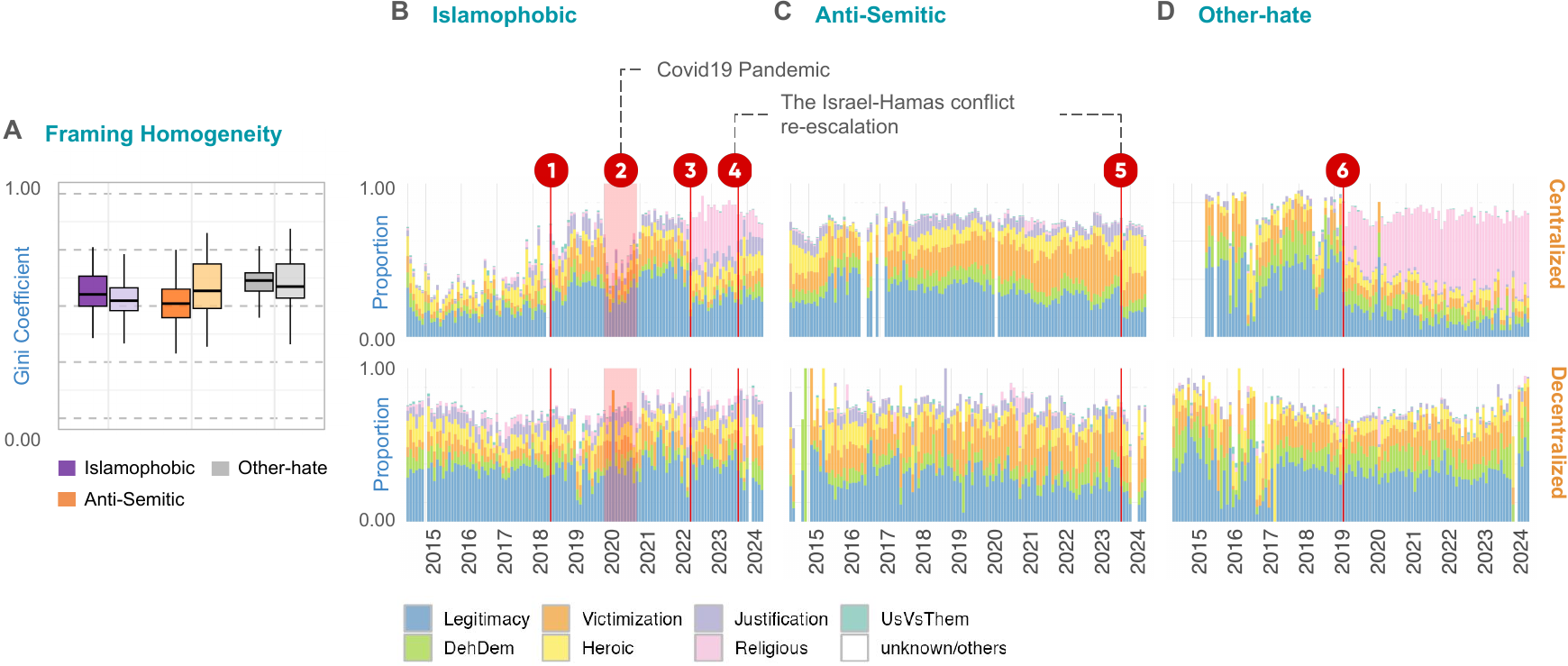}
    \caption{(A) Framing homogeneity by participation centralization across hate group ideologies (shades: darker = centralized; lighter = decentralized). Centralized Islamophobic groups exhibit greater framing homogeneity than their decentralized counterparts. In contrast, centralized Anti-Semitic groups display more diverse framings. The difference between centralized and decentralized Other-hate groups is not statistically significant. (B-D) Proportions of posts by narrative framing in Islamophobic, Anti-Semitic, and Other-hate groups by participation centralization. Markers numbered  \raisebox{0.2ex}{\redcircled{\scriptsize 1}} to \raisebox{0.2ex}{\redcircled{\scriptsize 6}} indicate the key events referenced in this results section. Visualizations demonstrate significant distinctions linked to participation structure and group ideology in terms of framing diversity. Note: Although our taxonomy includes eight narrative framing types, no post in the final dataset had \textit{Imminent War/Crisis} as its top framing, so it is excluded from the analysis.}
    \label{fig:frames_overtime}
\end{figure*}


\subsubsection{Narrative Framing Shifts and Homogeneity.}
Fig. \ref{fig:frames_overtime}A shows overall comparisons of framing homogeneity across ideologies and participation structures (see also Fig. \ref{fig:content_homogeneity_overtime}A for over-time Gini coefficient comparisons). Across ideologies, framing homogeneity is significantly higher in Other-hate groups (Gini coefficient of the framing distribution: $median = 0.60$, $IQR = 0.10$) compared to Islamophobic groups ($median = 0.54$, $IQR = 0.11$) and Anti-Semitic groups ($median = 0.53$, $IQR = 0.17$). Islamophobic and Anti-Semitic groups show no significant difference. See Appendix ``Additional Analysis Notes" for statistical test details.

Fig. \ref{fig:frames_overtime}B-D tracks narrative framing over time in centralized versus decentralized groups across ideologies. In centralized Islamophobic groups (Fig. \ref{fig:frames_overtime}B-top), although framing dominance was not clear before mid-2018 (\raisebox{0.2ex}{\redcircled{\scriptsize 1}}) and during COVID-19 pandemic (\raisebox{0.2ex}{\redcircled{\scriptsize 2}}), we can observe that before mid-2022 (\raisebox{0.2ex}{\redcircled{\scriptsize 3}}), \textit{Legitimacy} dominated among the identified framings. 
An example post\footnote{These examples were paraphrased using ChatGPT to protect user privacy while preserving the narrative strategy and expressions of hate. See Appendix ``Paraphrasing Narrative Examples'' for an example of the prompt and more information.} from this period demonstrates \textit{Legitimacy} framing, with the framing cues highlighted:\footnote{Although the examples in this paper contain hate speech, note that not every post in these hate groups constitutes hate speech or explicit hate speech.}

\begin{quote}
\small
\itshape
\hlsoftorange{I am an Ex-Muslim. Jesus appeared to me in a vivid dream,} which is why I became a Christian....
Jesus is not limited like Allah. Allah can only speak Arabic and understand Arabic. Allah is powerless and cannot transform into a human, a mountain, a lion, or a deer… or anything else. Why is the Islamic God like an Arab Mafia Boss instead of an unlimited God who doesn’t mind being drawn?.... 
\end{quote}

\noindent After mid-2022 (\raisebox{0.2ex}{\redcircled{\scriptsize 3}}), these groups primarily used \textit{Religious} (see an example below) and \textit{Legitimacy} frames, resulting in lower overall homogeneity in the decade (average monthly Gini before July 2022: $0.60$; after: $0.52$; also see Fig. \ref{fig:content_overtime}B).

\noindent \textit{Religious:}
\begin{quote}
\small
\itshape
.... \hlsoftorange{The Bible says, ``Do not allow a witch to live."} If a Christian mistakenly kills a Muslim, they are killing a Pagan, as Muslims are considered enemies of the Holy Spirit. But if a Muslim mistakenly kills a Christian in Christ, they have offended God, Jesus, and the Holy Spirit. \hlsoftorange{Christiandom is the way, as stated in John 14:6.}
\end{quote}

Decentralized Islamophobic groups (Fig. \ref{fig:frames_overtime}B-bottom) similarly presented \textit{Legitimacy} as the predominant frame, yet they exhibited lower homogeneity ($median = 0.52$, $IQR = 0.10$) than centralized ones ($median = 0.55$, $IQR = 0.13$; $p = 2.05 \times 10\textsuperscript{-4}$, $CI = [-0.05, -0.01]$; Fig. \ref{fig:frames_overtime}A), as their discourse contained a higher proportion of alternative framings beyond the dominant theme. This finding suggests that participation centralization is positively associated with more homogeneous framing among Islamophobic groups.




Conversely, centralized Anti-Semitic groups shows significantly lower framing homogeneity ($median = 0.51$, $IQR = 0.12$) than decentralized ones ($median = 0.57$, $IQR = 0.20$; $p < 10^{-6}$, $CI = [0.04, 0.07]$; Fig. \ref{fig:frames_overtime}A), as they often employed multiple frames (e.g., \textit{Victimization} alongside \textit{Legitimacy} (see examples below)), whereas decentralized groups mostly focused on \textit{Legitimacy} (Fig. \ref{fig:frames_overtime}C). These patterns suggest that greater participation centralization is associated with increased narrative variety among Anti-Semitic groups.

\noindent \textit{Legitimacy:}
\begin{quote}
\small
\itshape
\hlsoftorange{Ariel Sharon:} ``We, the Jewish People, control America, and Americans are fully aware of it!'' This video is likely to be targeted by the ADL, so share it before it disappears! Post it on Facebook and subscribe to her channel: [Youtube link]. 

\hlsoftorange{The Jewish Institute for National Security Affairs (JINSA) advisory board has included prominent figures such as Michael Ledeen, Richard Perle,} ....
\end{quote}

\begin{quote}
\small
[
Notes: Ariel Sharon is a prominent Israeli military and political leader, Michael Ledeen is an American foreign policy analyst, and Richard Perle is a former U.S. Assistant Secretary of Defense. The claim ``We, the Jewish People, control America....'' reflects a common antisemitic conspiracy portraying Jewish people as secretly controlling governments or media, a narrative that has historically fueled antisemitism.]
\end{quote}

\noindent \textit{Victimization:}
\begin{quote}
\small
\itshape
What does ``antisemitism'' really mean?\hlsoftorange{ It’s simply a scheme invented by Jews to trick and take advantage of Gentiles!}
\end{quote}

The Israel-Hamas conflict escalation on October 2023 coincided with a major narrative shift in both Islamophobic (\raisebox{0.2ex}{\redcircled{\scriptsize 4}}) (Fig. \ref{fig:frames_overtime}B) and Anti-Semitic groups (\raisebox{0.2ex}{\redcircled{\scriptsize 5}}) (\ref{fig:frames_overtime}C). Regardless of centralization, these groups adopted more diverse framings. For instance, in centralized Anti-Semitic groups (Fig. \ref{fig:frames_overtime}C-top \raisebox{0.2ex}{\redcircled{\scriptsize 5}}), the proportion of previously minor \textit{Heroic} framing (see an example below) increased, while the proportion of dominant framing, \textit{Legitimacy} declined, leading to a more balanced narrative landscape. 

\noindent \textit{Heroic:}
\begin{quote}
\small
\itshape
NO ONE supports Israehell, Jews, or Zionist yahudi lahnatullah. \hlsoftorange{Tom Hanks wearing a Palestinian scarf—just watch, it will appear everywhere.}
\end{quote}

Other-hate groups show similar levels of framing homogeneity between centralized ($median = 0.61$, $IQR = 0.09$) and decentralized structures ($median = 0.59$, $IQR = 0.15$; $p = 0.21$, $CI = [-0.05, -0.01]$; Fig. \ref{fig:frames_overtime}A). However, after mid-2019 (\raisebox{0.2ex}{\redcircled{\scriptsize 6}}), centralized groups (Fig. \ref{fig:frames_overtime}D-top) shifted from \textit{Legitimacy} to \textit{Religious} framing (see examples below), whereas decentralized groups (Fig. \ref{fig:frames_overtime}D-bottom) consistently emphasized \textit{Legitimacy} framing throughout.

\noindent \textit{Legitimacy:}
\begin{quote}
\small
\itshape
\hlsoftorange{In a clear attempt to violate American immigration laws,} pro-unauthorized immigration activists are meeting asylum seekers to instruct them on what to say to gain entry into the US. They will deceive and exploit loopholes to obtain asylum. We must shut the border to all immigrants now.
\end{quote}

\noindent \textit{Religious:}
\begin{quote}
\small
\itshape
If you think you’re already ``saved,'' then what’s Judgement Day for? \hlsoftorange{Matthew 24:13 (KJV): ``But anyone who keeps going until the end will be saved."} Ask your pastor who eats pork! Christianity will get you killed by Yah.

\end{quote}

\subsubsection{Topic Shifts and Homogeneity.}
To provide a more robust analysis of content homogeneity, we examined topic distributions across structures and ideologies. The pattern mirrors narrative framing: centralized Islamophobic groups focus on narrower, more consistent topics, while decentralized groups cover a broader range; in Anti-Semitic groups, centralization corresponds with greater topical diversity. Detailed on topic shifts and homogeneity across structures and ideologies are reported in the Appendix ``Topic Shifts and Homogeneity."

\subsection{Participation Structures and The Connection Patterns Among Hate Groups (RQ4)}

\paragraph{Summary of Findings:} \textit{There is no significant difference in affiliation tendencies between centralized and decentralized groups within each ideology. However, we found that Islamophobic groups are significantly more homophilic than anti-Semitic groups, which show a more balanced affiliation between groups of the same and different ideologies. }
\vspace{3pt}

\begin{figure}[h]
    \centering
    \includegraphics[width=0.83\columnwidth]{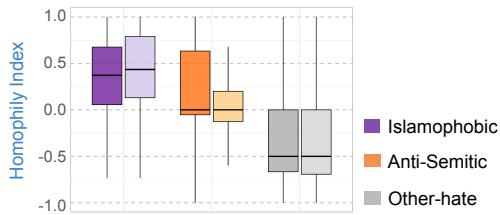} 
    \caption{Homophily across group ideologies and structures (shades: darker = centralized; lighter = decentralized). Islamophobic groups are more homophilic than Anti-Semitic and Other-hate groups; Other-hate groups are heterophilic. 
    }
    \label{fig:homophily}
\end{figure}

Beyond within-group dynamics, we explored how participation structures relate to cross-group engagement, which might reflect patterns of information diffusion. 
Our findings are shown in Fig. \ref{fig:homophily}. The statistical tests (see Appendix ``Additional Analysis Notes") suggest no significant link between participation structure and affiliation with groups of similar ideology. However, we found significant differences across hate group ideologies. Specifically, we found that Islamophobic groups in general are significantly more homophilic (weighted Homophily Index: $median = 0.40$, $IQR = 0.65$) than Anti-Semitic groups ($p < 10\textsuperscript{-6}$, $CI$ for $median$ difference $= [0.33, 0.45]$), which exhibit a balanced affiliation between groups of the same and different ideologies ($median = 0.00$, $IQR = 0.60$; Fig. \ref{fig:homophily}). This suggests that narratives from Anti-Semitic groups are more likely to spread beyond groups with the same ideology, while those from Islamophobic groups are more likely to remain within like-minded ones. Previously, in the regression analysis we observed that in Anti-Semitic groups, homophily is negatively associated with future engagement ($\beta_{homophily} = -0.12$, $p = 0.01$, $CI = [-0.21, -0.03]$; Fig. \ref{fig:regression_analysis}). This pattern may help explain why anti-Semitic groups appear less homophilic than Islamophobic groups, though further analysis is needed.

Finally, we found that Other-hate groups are heterophilic ($median = -0.50$, $IQR = 0.67$), with a homophily index significantly lower than both Islamophobic ($p < 10\textsuperscript{-6}$, $CI = [0.78, 0.94]$) and Anti-Semitic ($p < 10\textsuperscript{-6}$, $CI = [0.40, 0.50]$) groups, meaning that Islamophobic and Anti-Semitic groups tend to be more isolated compared to groups with ideologies other than these two.

\section{Discussion}
Using a decade of Facebook data on anti-Semitic and Islamophobic communities, this study examines how participation structures relate to engagement patterns in online hate groups. Previous works suggest that both centralized and decentralized communities can generate high levels of interaction \cite{kasimov2025decentralized, poole2021tactical}. In this study, across hate group ideologies related to Israel-Palestine conflict, we find that participation centralization is more effective in sustaining engagement. Central actors—those who dominate content creation and amplification—are linked with sustained community activity. 

This pattern may be explained by the relative complexity of sustaining activity in decentralized communities, which depends on the alignment of multiple factors, including prompt user engagement, individual autonomy, acceptance of unconventional or extreme strategies such as accelerationist tactics, and the maintenance of anonymity \cite{kasimov2025decentralized}. By contrast, centralized communities are supported by well-defined hierarchies and leadership structures, which facilitate mobilization and allow a limited number of central actors to consistently sustain community activity \cite{kasimov2025decentralized}. 

This finding aligns with the club goods framework, which emphasizes how organization fosters group cohesion \cite{iannaccone1992sacrifice, berman2008religion}. It also corresponds with studies showing that strong leadership supports group endurance \cite{freilich2009critical} and that the loss of key members often precipitates group dissolution \cite{jones2008terrorist}.


To further explore how participation structures relate to engagement patterns, we assessed whether central actors correspond to variation in narrative diversity. As noted earlier \cite{kasimov2025decentralized}, centralized communities rely on opinion leaders, while decentralized communities depend on dispersed, autonomous participation, and tolerance for unorthodox tactics. Thus, we expected that centralized groups would produce more uniform content, whereas decentralized communities would generate more diverse messages. Our findings partially support this: centralized Islamophobic groups showed more uniform content. This pattern is consistent with prior work suggesting key actors in hybrid media spaces can coordinate messaging effectively \cite{postill2018populism}, whereas decentralized digital networks allow for more diverse participation and messaging \cite{ganesh2018ungovernability}. However, centralized anti-Semitic groups showed broader narrative diversity, contrary to the prediction.

This divergence likely reflects the distinct historical and political contexts of these ideologies. Anti-Semitic narratives have developed amid complex cultural and political pressures \cite{samuels2018literature}, where leaders may pursue differing agendas with limited coordination. In contrast, Islamophobic groups often center on immediate political issues \cite{bertran2018islamophobia}, resulting in more focused and consistent leadership aligned with specific real-time events.

Another possible explanation for the consistency in Islamophobic messaging is that it likely reflects deliberate coordination among leaders in offline activism. Groups such as the Identitarian Movement, along with European populist radical right parties, exemplify centralized leadership, transnational networks, and formal structures that propagate anti-Muslim rhetoric echoed in mainstream media and politics \cite{zuquete2018identitarians, brubaker2017between, mudde2007populist, norris2019cultural, mudde2019farright, goodwin2011new}. This pattern contrasts with anti-Semitic communities, which are often covert and fragmented, operating in fringe or encrypted spaces that emphasize anonymity and foster isolated acts rather than coordinated mobilization \cite{gwu2022antisemitism, sanborn2020confronting}.

These findings have implications for countering online Islamophobia and anti-Semitism. In Islamophobic groups, targeting centralized leadership may reduce hateful narratives. Although removing key figures can provoke short-term backlash, it typically lowers engagement over time \cite{thomas2023disrupting}. Anti-Semitic groups, by contrast, operate through dispersed networks, requiring broader strategies for both leaders and participants. However, even with less coordinated leaders, centralization remains important, as it is associated with higher engagement and greater content diversity, which in turn predicts engagement.

We finally examined how participation structures relate to intergroup engagement, assessing whether narratives circulate mainly within ideologically similar groups (homophily) or across ideologies (heterophily). Centralization did not predict homophily, but ideological patterns emerged: Islamophobic groups clustered tightly, while anti-Semitic groups maintained a mix of in- and out-group ties. 

These patterns carry distinct risks. Tight clustering in Islamophobic groups may create echo chambers, limiting exposure to alternative viewpoints and enabling hateful content escalation \cite{bakshy2015exposure, cinelli2021echo}. In contrast, anti-Semitic groups’ broader connections may reduce insularity but can backfire, strengthening preexisting biases through hostile cross-group interactions \cite{bail2018exposure, kumar2018community}. The stronger homophily in Islamophobic groups suggests interventions may be most effective within ideologically aligned communities, such as nudging users toward alternative perspectives \cite{currin2022depolarization, pal2023depolarization, celadin2024promoting}. For anti-Semitic groups, which span ideological lines, interventions should target potentially problematic bridging points, monitoring cross-group content that may provoke backlash.

\paragraph{Limitations and Future Work.} This study has several limitations. Given the complexity of hate group categorization, we adopted a broad definition of Anti-Semitic and Islamophobic sentiment, encompassing hate directed at Jewish and Islamic communities as well as individuals associated with Israel and Palestine. While this captures the intertwined nature of these narratives, it limits the ability to distinguish between racial/religious and geopolitical hate. 

Additional considerations relate to the scope of our study. Our data focus on Anti-Semitic and Islamophobic communities, which limits generalizability to other hate groups or socio-political contexts. Although an Other-hate category was included, it is heterogeneous, encompassing subgroups such as white supremacy and anti-feminism, and is further limited to groups related to the Israel–Hamas conflict. Banned Facebook groups were excluded, and keyword-based searches may have missed some relevant groups, potentially introducing bias. Textual analyses were also limited to English-language posts (70.28\% of text-based posts). Future research could address these limitations by examining other hate groups across diverse ideological and linguistic contexts.

\section{Acknowledgments}
The authors would like to acknowledge support from AFOSR, ONR, Minerva, NSF \#2318461, and Pitt Cyber Institute's PCAG awards. The research was partly supported by Pitt's CRC resources. We gratefully acknowledge Dr. Deborah Wheeler, Dr. Andrew Miller, and the students of the United States Naval Academy for their valuable early input on this research. The authors also thank Wanhao Yu for contributions to the dataset and technical consultation, the annotators for their valuable work, and the anonymous reviewers for their constructive feedback. Any opinions, findings, and conclusions or recommendations expressed in this material do not necessarily reflect the views of the funding sources.




\subsection{Paper Checklist}

\begin{enumerate}

\item For most authors...
\begin{enumerate}
    \item  Would answering this research question advance science without violating social contracts, such as violating privacy norms, perpetuating unfair profiling, exacerbating the socio-economic divide, or implying disrespect to societies or cultures?
    \answerYes{Yes, see ``Ethical Considerations.''}
  \item Do your main claims in the abstract and introduction accurately reflect the paper's contributions and scope?
    \answerYes{Yes, our main claims in the abstract and introduction summarize what we did and what we found, as described in the ``Methodology'' and ``Results.''}
   \item Do you clarify how the proposed methodological approach is appropriate for the claims made? 
    \answerYes{We justified our methodology in ``Methodology'' and provided detailed explanations in the Appendix ``Methodological Choices.'' We also discussed possible limitations in ``Limitations and Future Work'' to clarify considerations for interpreting the findings.}
   \item Do you clarify what are possible artifacts in the data used, given population-specific distributions?
    \answerYes{Yes, see ``Limitations and Future Work.''}
  \item Did you describe the limitations of your work?
    \answerYes{Yes, see ``Limitations and Future Work.''}
  \item Did you discuss any potential negative societal impacts of your work?
    \answerYes{Yes, see ``Ethical Considerations.''}
      \item Did you discuss any potential misuse of your work?
    \answerYes{Yes, see ``Ethical Considerations.''}
    \item Did you describe steps taken to prevent or mitigate potential negative outcomes of the research, such as data and model documentation, data anonymization, responsible release, access control, and the reproducibility of findings?
    \answerYes{Yes, see ``Ethical Considerations.''}
  \item Have you read the ethics review guidelines and ensured that your paper conforms to them?
    \answerYes{Yes.}
\end{enumerate}

\item Additionally, if your study involves hypotheses testing...
\begin{enumerate}
  \item Did you clearly state the assumptions underlying all theoretical results?
    \textcolor{gray}{NA}
  \item Have you provided justifications for all theoretical results?
    \textcolor{gray}{NA}
  \item Did you discuss competing hypotheses or theories that might challenge or complement your theoretical results?
    \textcolor{gray}{NA}
  \item Have you considered alternative mechanisms or explanations that might account for the same outcomes observed in your study?
    \textcolor{gray}{NA}
  \item Did you address potential biases or limitations in your theoretical framework?
    \textcolor{gray}{NA}
  \item Have you related your theoretical results to the existing literature in social science?
    \textcolor{gray}{NA}
  \item Did you discuss the implications of your theoretical results for policy, practice, or further research in the social science domain?
    \textcolor{gray}{NA}
\end{enumerate}

\item Additionally, if you are including theoretical proofs...
\begin{enumerate}
  \item Did you state the full set of assumptions of all theoretical results?
    \textcolor{gray}{NA}
	\item Did you include complete proofs of all theoretical results?
    \textcolor{gray}{NA}
\end{enumerate}

\item Additionally, if you ran machine learning experiments...
\begin{enumerate}
  \item Did you include the code, data, and instructions needed to reproduce the main experimental results (either in the supplemental material or as a URL)?\newpage
    \answerYes{While we are unable to share the raw content data to protect user privacy and comply with Meta’s platform policies, we provide the model, code, and aggregated statistics, which include group identifiers, to support transparency and reproducibility. The URL to the resources is available in the Appendix ``Materials and Code Availability.'' 
    }
  \item Did you specify all the training details (e.g., data splits, hyperparameters, how they were chosen)?
    \answerYes{Yes, see Appendix ``Extremist Narrative Framing Classification.''}
     \item Did you report error bars (e.g., with respect to the random seed after running experiments multiple times)?
    \answerNo{We did not report error bars, as the experiments were conducted without specifying random seeds. 
    }
	\item Did you include the total amount of compute and the type of resources used (e.g., type of GPUs, internal cluster, or cloud provider)?
    \answerYes{We used the University of Pittsburgh Center for Research Computing and Data (CRC). Details on the computing hardware are available at \url{https://crc.pitt.edu/overview-crc-services/computing-hardware}.}
     \item Do you justify how the proposed evaluation is sufficient and appropriate to the claims made? 
    \answerYes{Yes, and it can be found in Methodology ``Machine classification of narrative frames.''}
     \item Do you discuss what is ``the cost`` of misclassification and fault (in)tolerance?
    \answerYes{Yes, a discussion of potential misclassification costs and fault tolerance is available in the Appendix ``Extremist Narrative Framing Classification: Comparison of Narrative Classifiers.''}
  
\end{enumerate}

\item Additionally, if you are using existing assets (e.g., code, data, models) or curating/releasing new assets, \textbf{without compromising anonymity}...
\begin{enumerate}
  \item If your work uses existing assets, did you cite the creators?
    \answerYes{Yes, we cited the creators of the assets in ``Methodology'' and ``Appendix.''}
  \item Did you mention the license of the assets?
    \answerYes{Yes, we mentioned the licenses of the external assets; see Ethical Considerations ``Data Collection and Management'' and Appendix ``Asset Licenses.''}
  \item Did you include any new assets in the supplemental material or as a URL?
    \answerYes{Yes, they are available in the Appendix ``Materials and Code Availability.''}
  \item Did you discuss whether and how consent was obtained from people whose data you're using/curating?
    \answerYes{Yes, see Ethical Considerations ``Data Collection and Management.''}
  \item Did you discuss whether the data you are using/curating contains personally identifiable information or offensive content?
    \answerYes{Yes, we discuss privacy concerns in the ``Ethical Considerations'' and provide a warning about offensive content at the beginning of this paper, which is also addressed in more detail in the ``Ethical Considerations.''}
\item If you are curating or releasing new datasets, did you discuss how you intend to make your datasets FAIR (see FORCE11 (2020))?
 \answerNA{NA}
\item If you are curating or releasing new datasets, did you create a Datasheet for the Dataset (see Gebru et al.
(2021))? 
\answerNA{NA}
\end{enumerate}

\item Additionally, if you used crowdsourcing or conducted research with human subjects, \textbf{without compromising anonymity}...
\begin{enumerate}
  \item Did you include the full text of instructions given to participants and screenshots?
    \textcolor{gray}{NA}
  \item Did you describe any potential participant risks, with mentions of Institutional Review Board (IRB) approvals?
    \textcolor{gray}{NA}
  \item Did you include the estimated hourly wage paid to participants and the total amount spent on participant compensation?
    \textcolor{gray}{NA}
   \item Did you discuss how data is stored, shared, and deidentified?
   \textcolor{gray}{NA}
\end{enumerate}

\end{enumerate}

\section{Ethical Considerations 
}

\paragraph{Data Collection and Management.}
Data for this study was collected using Meta's CrowdTangle API, which provided access to public content from Facebook Pages and Groups at the time of collection. However, this API has since been deprecated and is no longer publicly available. In line with Meta's platform policies and terms of service, we are unable to share the raw content data directly. We do, however, provide aggregated statistics and group identifiers to support transparency and reproducibility.


\paragraph{User Anonymity and Privacy.}
There was no direct contact with Facebook users, and no private data were obtained.
Results in this study are presented anonymously or in aggregate, with no user-specific information disclosed. 
A few Facebook posts are included as examples; these have been paraphrased using ChatGPT to prevent tracing back to the original users (see Appendix ``Paraphrasing Narrative Examples"), and we verified through search that the user information could not be easily recovered.

\paragraph{Socio-Political Impact.} The narrative examples presented in this paper include hateful content, which may provoke anxiety, anger, or social polarization. A warning at the beginning of the paper allows readers to engage with the material mindfully or discontinue reading if they prefer.

\paragraph{Annotation Complexity and Subjectivity.} To identify the hate groups analyzed in this study, graduate students reviewed posts for hate speech targeting Jews, Muslims, or ideologies such as White Supremacy. This process may introduce bias; to reduce the risk, an author without conflicts of interest conducted an additional review. For annotating hate narrative frames, we developed a detailed codebook with definitions, inclusion criteria, and dataset examples. Final labels were assigned by majority vote. We acknowledge that annotation is inherently subjective and prone to human error, and that different researchers may reach different conclusions despite consensus among trained annotators.

\paragraph{Potential Use/Abuse of Work.}
This study examines how online hate groups operate through participation structures and hate ideologies and highlights potential implications for addressing online hate. However, these findings could be misused by hate groups to bypass proposed interventions. We emphasize the need for continuous research and careful monitoring to improve intervention strategies and prevent potential misuse of the results.

\appendix
\section{Appendix}
\renewcommand{\thefigure}{A\arabic{figure}}
\renewcommand{\thetable}{A\arabic{table}}
\setcounter{figure}{0}
\setcounter{table}{0}

\subsection{Topic Shifts and Homogeneity}

\begin{figure*}[ht]
    \centering
    \includegraphics[width=1.0\textwidth]{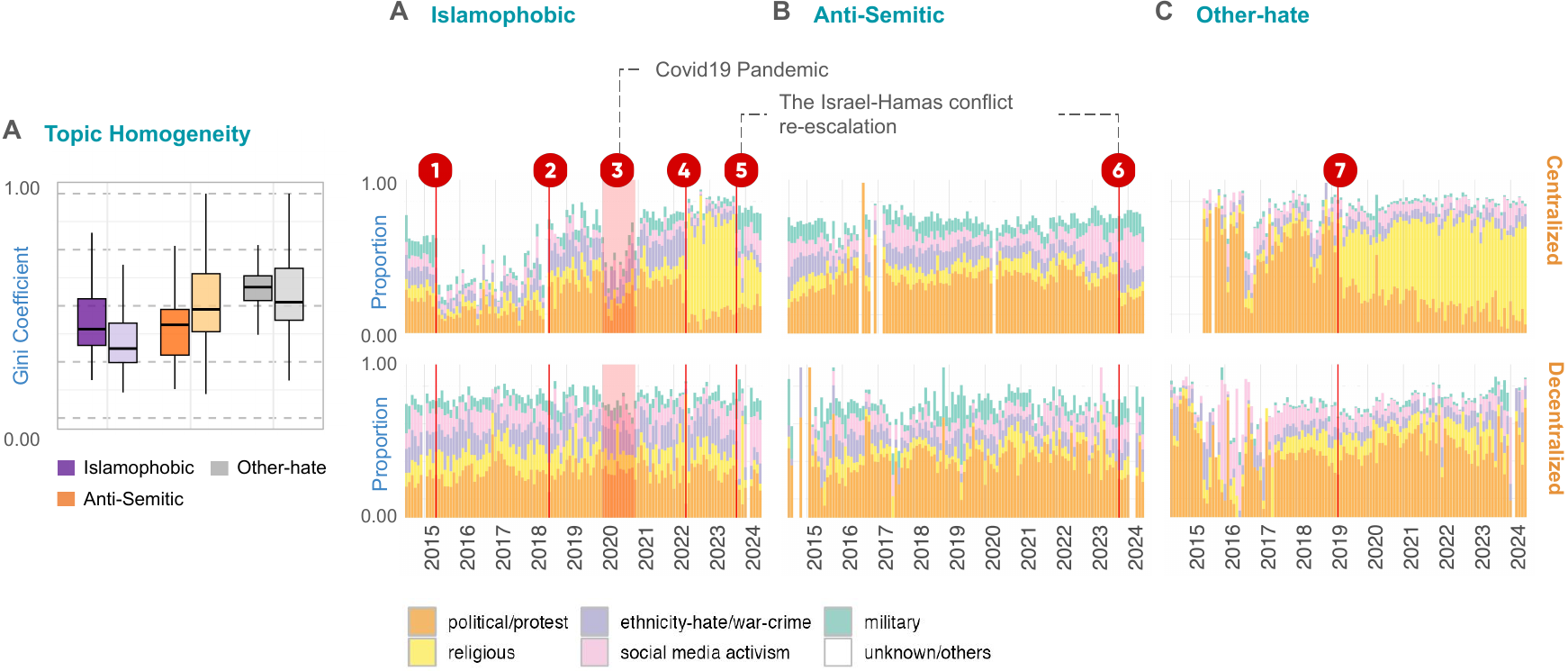} 
    \caption{Proportions of posts by discussion topic in (A) Islamophobic, (B) Anti-Semitic, and (C) Other-hate groups by participation centralization. Markers numbered  \raisebox{0.2ex}{\redcircled{\scriptsize 1}} to \raisebox{0.2ex}{\redcircled{\scriptsize 7}} indicate the key events referenced in this results section. Visualizations demonstrate significant distinctions linked to participation structure and group ideology in terms of topic diversity.}
    \label{fig:topics_overtime}
\end{figure*}

Fig. \ref{fig:topics_overtime}A presents overall comparisons of topic homogeneity across ideologies and participation structures (see also Fig. \ref{fig:content_homogeneity_overtime}B for over-time Gini coefficient comparisons).
Overall, Other-hate groups had significantly higher topic homogeneity (Gini coefficient of the topic distribution: $median = 0.56, IQR = 0.21$) compared to Islamophobic groups ($median = 0.36, IQR = 0.19$) and Anti-Semitic groups ($median = 0.44, IQR = 0.21$). Anti-Semitic groups had significantly higher topic homogeneity than Islamophobic groups. See Appendix: Additional Analysis Notes for statistical test details.

Fig. \ref{fig:topics_overtime}B-D compares topical trends over time between centralized and decentralized groups within each ideology. Among centralized Islamophobic groups (Fig. \ref{fig:topics_overtime}B-top), topical dominance was less evident between mid-2015 (\raisebox{0.2ex}{\redcircled{\scriptsize 1}}) and mid-2018 (\raisebox{0.2ex}{\redcircled{\scriptsize 2}}) as well as during the COVID-19 pandemic (\raisebox{0.2ex}{\redcircled{\scriptsize 3}}). However, among the identified topics, prior to mid-2022 (\raisebox{0.2ex}{\redcircled{\scriptsize 4}}), discussions on \textit{Political/Protest} topics emerged as the most prominent, for instance (with the topic cues highlighted): 

\begin{quote}
\small
\itshape
....\hlgreen{It’s time to boycott these racist Muslims. There’s a petition demanding that FIFA move the games out of Qatar, and we should all sign it right away. We also need to put together a list of what they own, make it public, and boycott everything connected to Qatar.} Thousands of migrant workers in Qatar die while earning wages comparable to slavery....
\end{quote}

\noindent Around mid-2022 (\raisebox{0.2ex}{\redcircled{\scriptsize 4}}), \textit{Religious} topics became predominant, for example: 

\begin{quote}
\small
\itshape
\hlgreen{Here are the facts about Muhammad: He was born and raised as a pagan.... The Quran and Hadith themselves admit he was a sinner. He never performed any miracles and had no knowledge of the future — and neither did Allah.} He was not a man of peace; instead, he built an army, killed people, captured and sold women, looted, and used terror....
\end{quote}

Like their centralized counterparts, decentralized Islamophobic groups (Fig. \ref{fig:topics_overtime}B-bottom) consistently focused on \textit{Political/Protest} topics. However, besides \textit{Political/Protest}, the decentralized ones also discussed a wider range of topics, such as \textit{Social Media Activism}
(see an example below). 

\begin{quote}
\small
\itshape
....The pro-Muslim movement is just like the Nazi machine.... We cannot keep sacrificing Jews to this Muslim, Nazi-like force that hides behind ``religion'' and ``freedom of speech'' while it terrorizes other faiths and violently strips people and their supporters of their freedoms.\hlgreen{ Please share your thoughts on the Brooklyn College page: [a Facebook link].}
\end{quote}

\noindent Overall, centralized Islamophobic groups had significantly higher topic homogeneity ($median = 0.40, IQR = 0.21$)  than their decentralized counterparts ($median = 0.31, IQR = 0.18$; $p < 10\textsuperscript{-6}, CI = [-0.11, -0.05]$; Fig. \ref{fig:topics_overtime}A), indicating a link between participation centralization and topical focus.

In Anti-Semitic groups (Fig. \ref{fig:topics_overtime}C), both centralized and decentralized groups also primarily discussed \textit{Political/Protest}, such as:

\begin{quote}
\small
\itshape
Thanks to Jew Cullen for pointing this out. \hlgreen{In response to the fires, Scotty from Marketing is making an authorized political ad for the Liberal Party, complete with upbeat music.} Absolutely unbelievable. [angry face][angry face][angry face]
\end{quote}

\noindent However, centralized groups engaged more frequently with additional topics (especially \textit{Social Media Activism}; see an example below), resulting in significantly lower topic homogeneity ($median = 0.42, IQR = 0.20$) compared to decentralized groups ($median = 0.49, IQR = 0.26$; $p < 10\textsuperscript{-6}, CI = [0.04, 0.10]$; Fig. \ref{fig:topics_overtime}A). 

\noindent \textit{Social Media Activism:}
\begin{quote}
\small
\itshape
\hlgreen{Stop the Jewish Criminal Protection Bill in the Senate now! H.Res.707 defends Jewish criminals and takes away Americans’ First Amendment rights!}
\end{quote}

\noindent This pattern contrasts directly with the pattern seen in Islamophobic groups, suggesting opposite relationships between participation centralization and topical diversity.

Similar to narrative framings, the escalation of the Israel-Hamas conflict on October 2023 also coincided with shifts in discourse topics across both centralized and decentralized Islamophobic (\raisebox{0.2ex}{\redcircled{\scriptsize 5}})
 and Anti-Semitic groups (\raisebox{0.2ex}{\redcircled{\scriptsize 6}}) (Fig. \ref{fig:topics_overtime}B and \ref{fig:topics_overtime}C). At this point of time, the most dominant topics declined in prominence, less prominent topics like \textit{Social Media Activism} gained traction, and the overall use of discourse topics became more balanced.

Among Other-hate groups, centralized discussions (Fig. \ref{fig:topics_overtime}D-top) focused predominantly on \textit{Political/Protest} until early 2019 (\raisebox{0.2ex}{\redcircled{\scriptsize 7}}), then shifted toward \textit{Religious} topics (see examples below). 

\noindent \textit{Political/Protest:}
\begin{quote}
\small
\itshape
\hlgreen{A Democrat Congresswoman of color is calling for lawlessness. Shouldn’t an elected official who swore to uphold the Constitution and promotes disorder in the streets face prosecution or removal?} I’m fed up with liberal women of color who think their gender and race excuse their hateful words. Their insecurity and immaturity don’t justify stirring up chaos. \hlgreen{All elected officials should be held accountable for encouraging lawlessness, no matter their race or gender. [Link]}
\end{quote}

\noindent \textit{Religious:}
\begin{quote}
\small
\itshape
Black people don’t bother learning or checking the facts before they celebrate, and then they call it tradition. \hlgreen{Read what the Bible actually says about tradition} if you’re not too lazy to read it!
\end{quote}

\noindent In contrast, decentralized Other-hate groups (Fig. \ref{fig:topics_overtime}D-bottom) maintained a consistent focus on \textit{Political/Protest}. Overall, bootstrapped Mann--Whitney \textit{U} test with the significance level adjusted for multiple comparisons shows that the level of topic homogeneity between centralized ($median = 0.58, IQR = 0.11$) and decentralized Other-hate groups ($median = 0.52, IQR = 0.23$) was similar ($p = 7.08 \times 10\textsuperscript{-3}, CI = [-0.09, -0.04]$; Fig. \ref{fig:topics_overtime}A). However, in centralized groups, the dominant topic shifted over time, whereas in decentralized groups, it remained more stable with no significant changes over the decade.

\subsection{Hate Group Distribution}
Table \ref{tab:group_summary} summarizes the distribution of hate groups analyzed in this study.

\renewcommand{\arraystretch}{1.3} 

\begin{table}[h!]
\centering\small
\caption{Summary of groups, users, posts, and monthly groups across hate group categories.}
\label{tab:group_summary}
\begin{tabular}{lcccc}
\hline
\textbf{Category} & \textbf{\#Groups} & \textbf{\#Monthly Groups} & \textbf{\#Users} & \textbf{\#Posts} \\
\hline
Islamophobic  & 5  & 444 & 9360  & 221025 \\
Anti-Semitic  & 14 & 887 & 15649 & 647873 \\
Other-hate    & 5  & 489 & 3633  & 126818 \\
\hline
\end{tabular}
\end{table}

\subsection{Methodological Choices}

\subsubsection{Choosing Gini Coefficient to Measure Participation Structure.} We chose to use the Gini coefficient to measure the concentration of group participation due to its simplicity and interpretability \cite{linRisingTidesRising2014}, with a direct connection with a graphical representation (Lorenz curve). Moreover, it has several advantages over alternative metrics. 

For example, degree centralization and power-law parameters are well-known metrics in studying networks. Degree concentration measures how unequal the degree distribution is in a network. It often focuses on the deviation of a single central node from others, making it sensitive to outliers. In contrast, Gini captures inequality across all nodes, not just in relation to the most central. Power-law parameter (typically denoted as $\alpha$) indicates the steepness of the power-law distribution in a network. It is often used to describe hub dominance. However, unlike Gini evaluating the entire distribution, it mainly captures the tail of the distribution, which focuses on the most connected nodes and ignores mid-range or low-degree nodes.

Shannon entropy is a widely used measure of uncertainty or diversity within a distribution. Compared with Gini, Shannon entropy is less sensitive to dominance and focuses on overall diversity, while the Gini coefficient is more capable of detecting disparities, making it better suited for identifying dominance in skewed distributions.

Simpson's diversity index is commonly used in ecology; it measures the probability that two individuals randomly selected from a sample belong to the same species. However, like entropy-based measures, Simpson's index is more focused on diversity rather than inequality. The Theil index is also an entropy-based measure of economic inequality. However, compared to the Gini coefficient, it is more complex to calculate and interpret.

The Herfindahl-Hirschman Index (HHI) is often used to measure market concentration. It is calculated by summing the squares of the market shares of all firms within an industry. The squaring of shares makes it overly sensitive to the presence of dominant members. In contrast, the Gini coefficient considers the entire distribution of member shares, offering a more balanced view of inequality.

\subsubsection{Inferring Narrative Framing and Topic.}
To infer narrative framing and topic for each Facebook post, we assign only the top-ranked framing and topic. Unlike news articles, which typically contain multiple paragraphs and extensive text, Facebook posts tend to be concise (median = 25 words). Given this brevity, we consider it more appropriate to categorize each post in our dataset with a single framing and single topic.

Note: While our taxonomy and model include eight narrative framing types, no post in our final dataset had \textit{Imminent War/Crisis} as its top-ranked framing. As a result, this framing type is not included in Figure \ref{fig:frames_overtime} or our analysis.

\subsubsection{Using Median as The Threshold for Dichotomizing Centralization.}
The distribution of Gini coefficient is bimodal, with a smaller second peak at the higher end, likely contributing to the sensitivity of the threshold choice. To explore this, we reanalyzed the data using the mean instead of the median as the classification threshold. The results were largely consistent, with only one of nine becoming statistically insignificant, indicating that our findings remain robust when the mean is used as an alternative threshold. However, the median is still preferred because it is less sensitive to outliers and better reflects the central tendency of the bimodal distribution. Using the median resulted in a more balanced classification of groups as decentralized and centralized, ensuring an even split between the two categories (Table \ref{tab:group_centralization}). In contrast, the mean can be influenced by extreme values, potentially leading to a less balanced and skewed classification, making the median a more stable and interpretable cutoff.

\renewcommand{\arraystretch}{1.3} 

\begin{table}[h!]
\centering
\small
\caption{Total number of groups within group ideologies based on monthly participation (each group-month is counted separately). Dichotomizing groups by the median Gini coefficient of participation centralization reveals a more even distribution between centralized and decentralized groups, compared to using the mean.}
\label{tab:group_centralization}
\begin{tabular}{lccc}
\hline
 & \textbf{Islamophobic} & \textbf{Anti-Semitic} & \textbf{Other-hate} \\
\hline
\multicolumn{4}{l}{\textit{Using median as the threshold}} \\
Centralized   & 212 & 476 & 212 \\
Decentralized & 232 & 411 & 277 \\
\hline
\multicolumn{4}{l}{\textit{Using mean as the threshold}} \\
Centralized   & 233 & 538 & 248 \\
Decentralized & 211 & 349 & 241 \\
\hline
\end{tabular}
\end{table}

\subsubsection{Using the Mann--Whitney \textit{U} Test for Comparing Distributions.}
While the Kolmogorov–Smirnov (KS) test can also be used to compare two non-normal distributions, it does not focus on whether one sample has higher or lower values at a central location. Instead, the KS test assesses whether the entire distribution differs in terms of shape, spread, or other characteristics. In this study, however, we are interested in determining whether one distribution tends to have higher or lower values than another. Therefore, the Mann--Whitney \textit{U} test is more appropriate, as it specifically compares the central tendency of the distributions and helps identify which one tends to have higher or lower values.

Details on the adjusted alphas and test results are provided in Appendix: Additional Analysis Notes.

\subsubsection{Using the Negative Binomial Model to Predict Next Month’s Engagement.}
The Negative Binomial model was chosen because it effectively handles count data with overdispersion \cite{ucla2025negative}. Unlike Poisson regression, which assumes equal mean and variance, our data show variance exceeding the mean \cite{ismail2007handling}. Thus, the Negative Binomial model provides a more suitable and robust analysis of engagement.

To assess the goodness of fit of the Negative Binomial models, we examined residual dispersion, calculated as the ratio of the Pearson chi-square or deviance statistic to its degrees of freedom \cite{cameron1990regression}. After outlier exclusions (which represent $\leq$ $1\%$ of the observations for each model), the dispersion values decreased to below 2, suggesting improved model fit. Importantly, excluding these observations did not alter the statistical and/or practical significance of the variables of interest (i.e., participation structure and content homogeneity), indicating that the models are robust to these outliers.

For the model specifications, see Appendix: Additional Analysis Notes: Regression Analysis.

\subsection{Additional Analysis Notes}
\subsubsection{Bonferroni Adjustment.} The adjusted alphas for the statistical tests are as follows.
\begin{itemize}
    \item For participation centralization, three comparisons between group ideologies resulted in an adjusted $\alpha = 0.017 \; (0.05/3)$.
    \item For content homogeneity, we combined Gini framing and Gini topic comparisons (12 total) due to their conceptual similarity, yielding an adjusted $\alpha = 0.004 \; (0.05/12)$.
    \item For homophily index comparisons, six comparisons resulted in an adjusted $\alpha = 0.008 \; (0.05/6)$.
\end{itemize}

\subsubsection{Participation Structure Across Group Ideologies.} 
The bootstrapped Mann--Whitney \textit{U} tests, with the significance level adjusted for multiple comparisons, confirm that the differences in centralization levels among categories are not statistically significant. Test results:
\begin{itemize}
    \item Other-hate vs.\ Islamophobic: $p = 0.77$, 95\% CI for median difference $= [-0.02, \, 0.04]$
    \item Other-hate vs.\ Anti-Semitic: $p = 0.28$, 95\% CI $= [0.02, \, 0.05]$
    \item Anti-Semitic vs.\ Islamophobic: $p = 0.24$, 95\% CI $= [-0.04, \, 0.01]$
\end{itemize}

\subsubsection{Framing Homogeneity Across Group Ideologies.} \noindent A bootstrapped Mann--Whitney \textit{U} test comparing the Gini coefficients of the framing distributions, with the significance level adjusted for multiple comparisons, shows that Other-hate groups have the highest framing homogeneity across group ideologies, while there is no significant difference between Anti-Semitic and Islamophobic groups. Test results:
\begin{itemize}
    \item Other-hate vs.\ Islamophobic: $p < 10^{-6}$, 95\% CI for median difference $= [-0.08, \, -0.05]$
    \item Other-hate vs.\ Anti-Semitic: $p < 10^{-6}$, 95\% CI $= [-0.08, \, -0.06]$
    \item Anti-Semitic vs.\ Islamophobic: $p = 0.69$, 95\% CI in homogeneity difference $= [-0.01, \, 0.02]$
\end{itemize}

\subsubsection{Topic Homogeneity Across Group Ideologies.}\noindent Other-hate groups exhibit the highest topic homogeneity, while Anti-Semitic groups show higher topic homogeneity than Islamophobic groups. Bootstrapped Mann--Whitney U tests comparing the Gini coefficients of the topic distributions, with the significance level adjusted for multiple comparisons, show:
\begin{itemize}
    \item Other-hate vs.\ Islamophobic: $p < 10^{-6}$, 95\% CI $= [-0.22, \, -0.17]$
    \item Other-hate vs.\ Anti-Semitic: $p < 10^{-6}$, 95\% CI $= [-0.13, \, -0.09]$
    \item Anti-Semitic vs.\ Islamophobic: $p < 10^{-6}$, 95\% CI $= [-0.13, \, -0.09]$
\end{itemize}

\subsubsection{Homophily by Participation Centralization Across Group Ideologies.} \noindent We found that centralized Islamophobic groups exhibit similar homophily (median $= 0.37$, IQR $= 0.62$) with the decentralized ones (median $= 0.43$, IQR $= 0.66$). Centralized Anti-Semitic groups show similar non-homophily (median $= 0.00$, IQR $= 0.68$) with the decentralized ones (median $= 0.00$, IQR $= 0.33$). In Other-hate groups, both centralized (median $= -0.50$, IQR $= 0.67$) and decentralized groups (median $= -0.50$, IQR $= 0.69$) exhibit similar levels of heterophily. Bootstrapped Mann--Whitney U tests comparing centralized vs.\ decentralized groups show:
\begin{itemize}
    \item Islamophobic groups: $p = 0.02$, 95\% CI $= [-0.04, \, 0.17]$
    \item Anti-Semitic groups: $p = 2.53 \times 10^{-4}$, 95\% CI $= [-0.11, \, 0.00]$
    \item Other-hate groups: $p = 0.98$, 95\% CI $= [-0.13, \, 0.10]$
\end{itemize}

\subsubsection{Monthly Engagement Distributions Across Group Ideologies.}

\begin{figure}[h!]
    \centering
    \includegraphics[width=0.8\columnwidth]{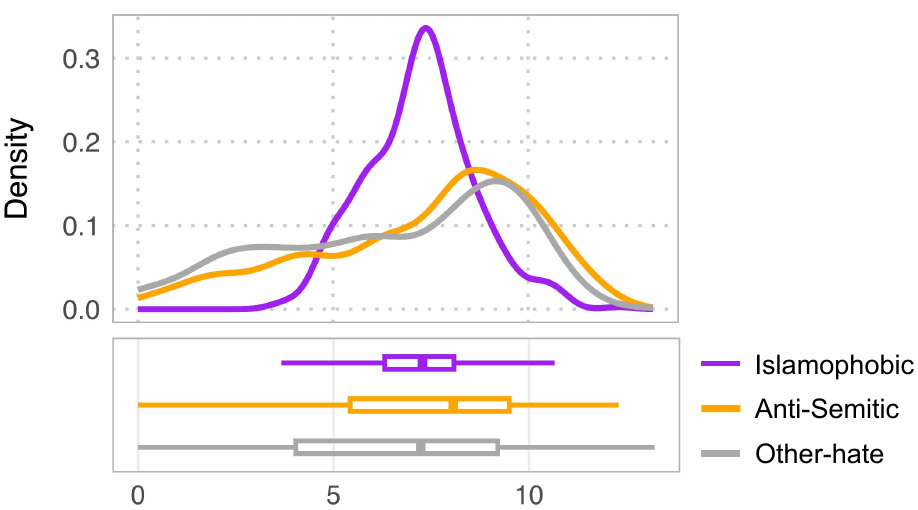}
    \caption{ Monthly total interactions (i.e., engagement) across group ideologies (log-transformed). Anti-Semitic groups tend to generate more engagement compared to the Islamophobic and Other-hate groups, with Islamophobic groups having no significant difference with Other-hate groups. }
    \label{fig:icwsm_engagement_dist}
\end{figure}

Overall, Anti-Semitic groups elicit higher engagement than Islamophobic and Other-hate groups, while engagement levels in Islamophobic and Other-hate groups do not differ significantly (Fig. \ref{fig:icwsm_engagement_dist}).

Total interactions per group (median and IQR):  
\begin{itemize}
    \item Islamophobic: median $= 1442$, IQR $= 2669$
    \item Anti-Semitic: median $= 3166$, IQR $= 13055$
    \item Other-hate: median $= 1382$, IQR $= 9828$
\end{itemize}

Bootstrapped Mann--Whitney \textit{U} tests comparing total interactions between groups show:
\begin{itemize}
    \item Anti-Semitic vs.\ Islamophobic: $p < 10^{-4}$, 95\% CI in total interactions difference $= [-2402, \, -1103]$
    \item Anti-Semitic vs.\ Other-hate: $p < 10^{-4}$, 95\% CI in total interactions difference $= [1820, \, 3485]$
    \item Islamophobic vs.\ Other-hate: $p = 0.2825$, 95\% CI in total interactions difference $= [-853, \, 2515]$
\end{itemize}

\subsubsection{Content Concentration Across Ideologies Overtime.}

\begin{figure}[h]
    \centering
    \includegraphics[width=0.85\columnwidth]{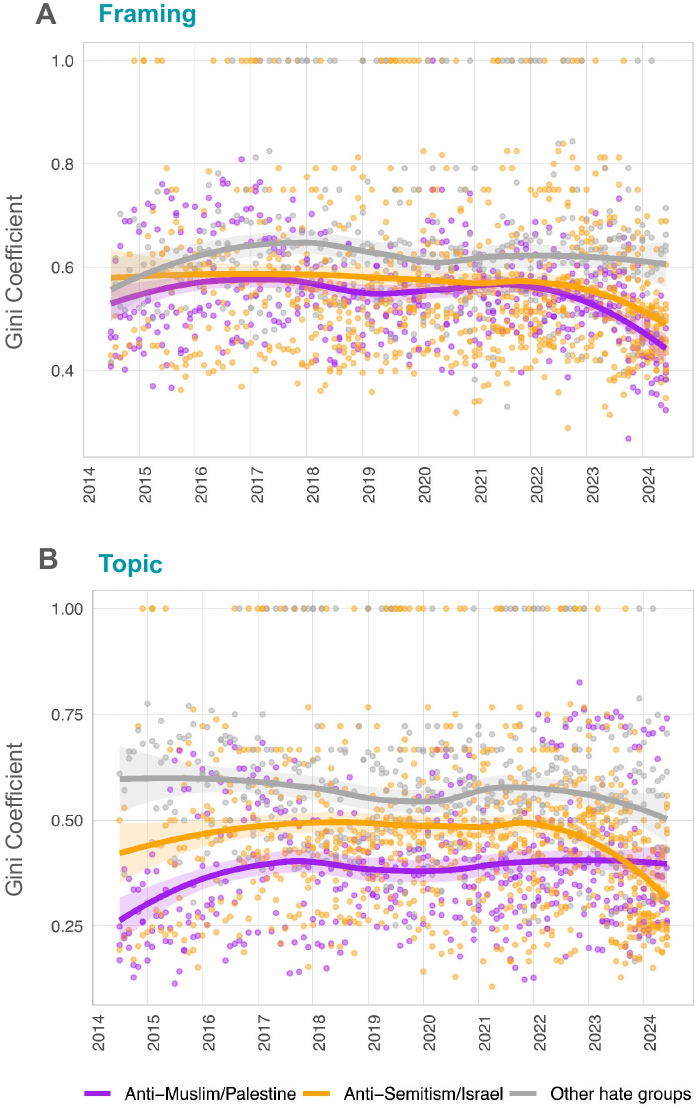}
    \caption{Concentration (Gini coefficients) of framing (A) and topic (B) across different group ideologies, with smoothed trend lines and 95\% CI ribbons. Figure (A) indicates that Other-hate groups had significantly higher framing concentration or homogeneity than Islamophobic and Anti-Semitic groups, while no significant difference was found between the latter two. Figure (B) shows that Other-hate groups also exhibited significantly greater topic concentration or homogeneity compared to both Islamophobic and Anti-Semitic groups, but Anti-Semitic groups displayed higher topic homogeneity than Islamophobic groups.}
    \label{fig:content_homogeneity_overtime}
\end{figure}

Fig. \ref{fig:content_homogeneity_overtime} shows content concentration across idelogies over the decade. Panel (A) indicates Other-hate groups had higher framing homogeneity than Islamophobic and Anti-Semitic groups, with no difference between the latter two. Panel (B) shows Other-hate groups also had greater topic homogeneity, while Anti-Semitic groups exceeded Islamophobic groups.

\subsubsection{Content Concentration Across Structures and Ideologies Overtime.}

\begin{figure*}[h]
    \centering
    \includegraphics[width=2\columnwidth]{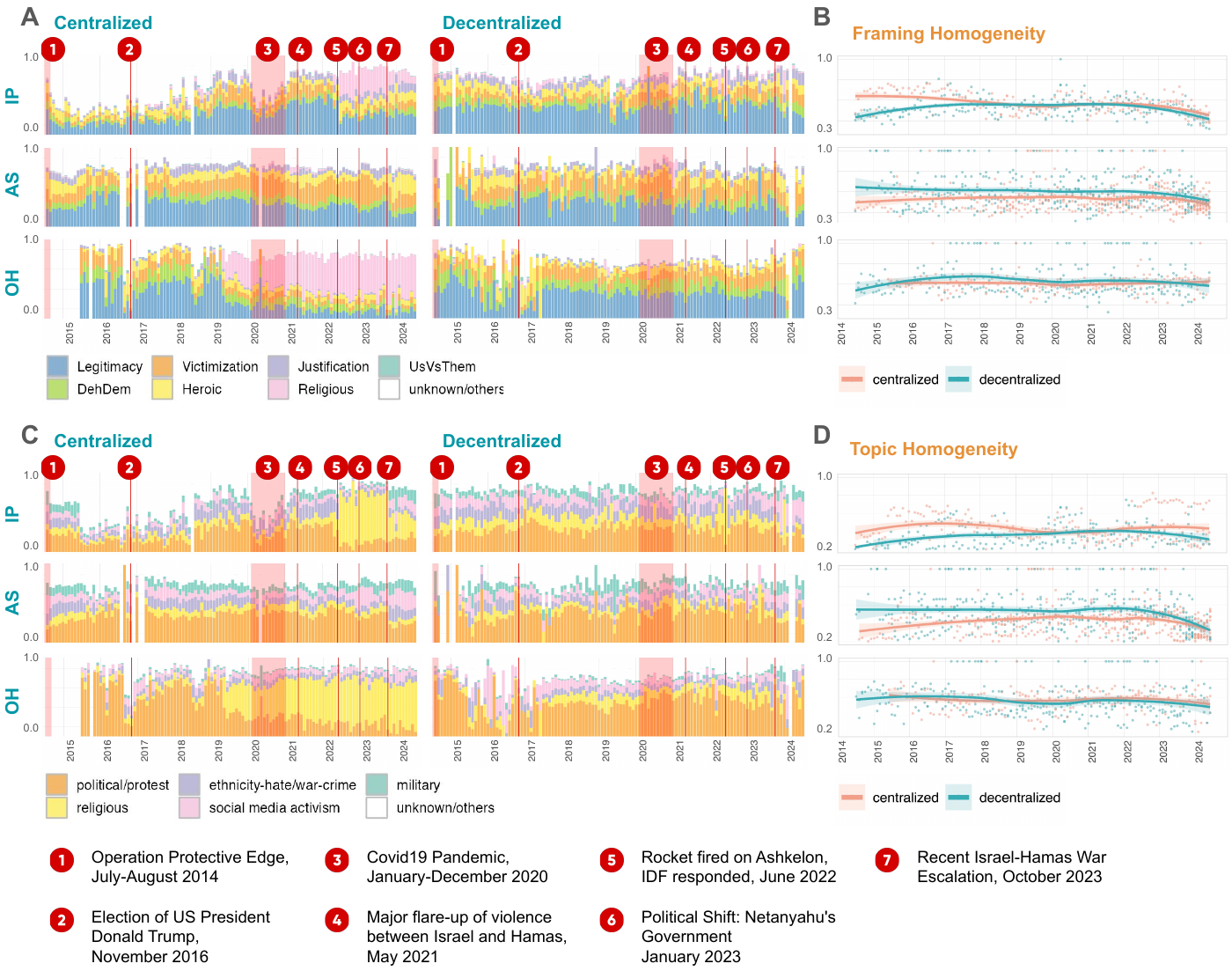}
    \caption{(A) Proportions of posts by narrative framing. (B) Framing homogeneity trends, comparing centralized and decentralized groups, with 95\% CI ribbons. (C) Proportions of posts by topical focus. (D) Topic homogeneity trends in centralized vs. decentralized groups, with smoothed trend lines and 95\% CI ribbons. Abbreviations: IP = Islamophobic; AS = Anti-Semitic; OH = Other-hate. Vertical red lines and shaded areas indicate key events. Visualizations demonstrate significant distinctions linked to participation structure and group ideology in terms of framing and topic diversity. 
    }
    \label{fig:content_overtime}
\end{figure*}

Fig. \ref{fig:content_overtime} presents content distribution over the decade, showing differences in framing and topic diversity across group ideologies and participation structures. Across the decade, the visualizations show differences in framing and topic diversity related to participation structure and group ideology.

\subsubsection{Regression Analysis.}
To examine the relationship between participation structure and engagement, we applied the Negative Binomial model to each hate group ideology and the overall data. 
This model regressed participation structure, content homogeneity, network attributes, framings, topics, and related factors to predict the following month’s engagement. Each standardized coefficient represents the expected change in next month’s engagement (i.e., total interactions) for a one-unit change in the predictor, holding other factors constant. Control variables included current engagement, median subscribers, and group age. Features were selected based on \textit{Akaike Information Criterion} (\textit{AIC}) improvements and ensuring the \textit{Variance Inflation Factor} (\textit{VIF}) remained below 4 to prevent multicollinearity.

\begin{figure*}[h]
    \centering
    \includegraphics[width=1.4\columnwidth]{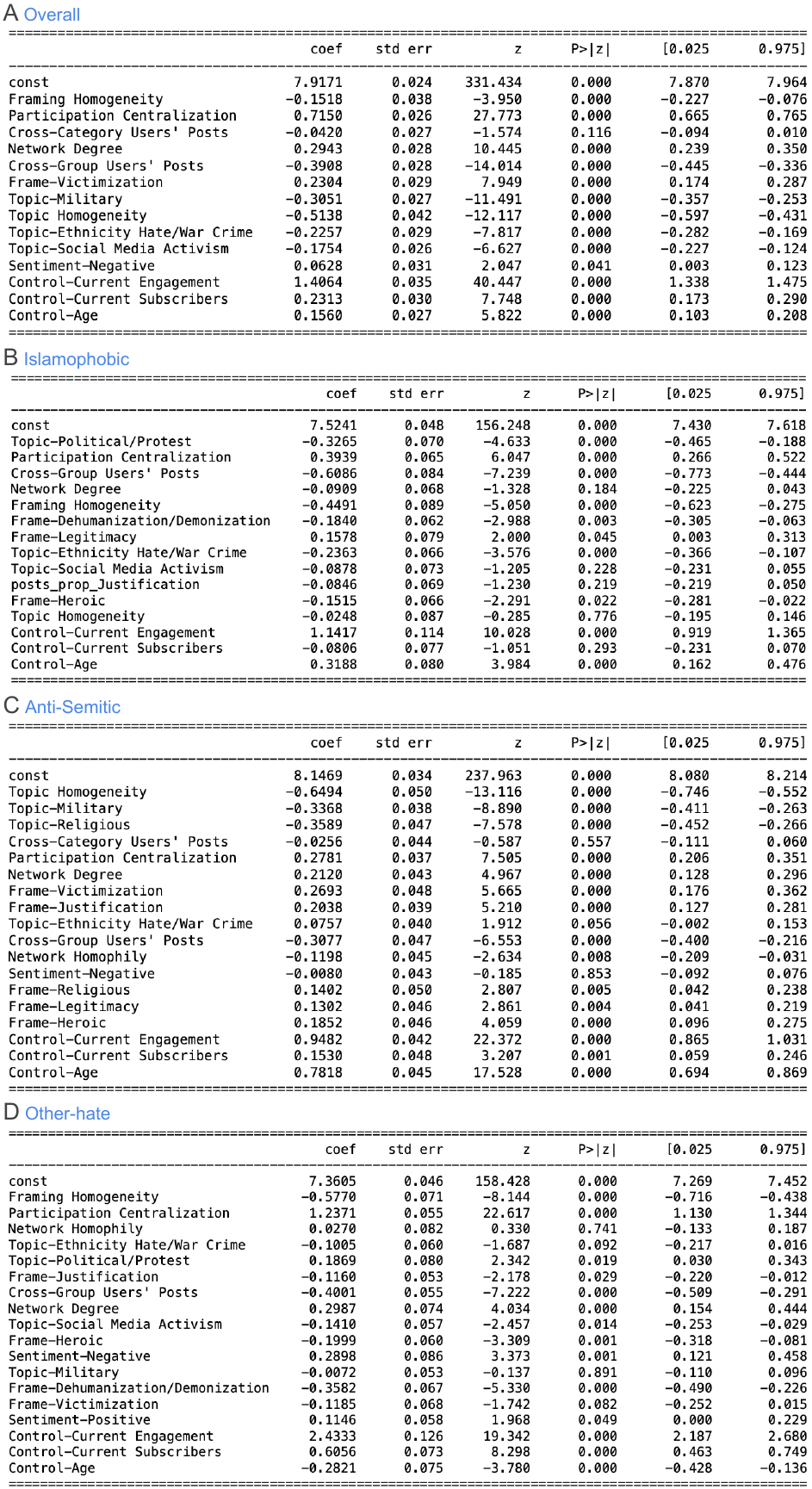}
    \caption{Regression results predicting next month's engagement across (A) the overall hate groups (Islamophobic, Anti-Semitic, and Other-hate groups); (B) Islamophobic groups; (C) Anti-Semitic groups, and (D) Other-hate groups.}
    \label{fig:complete_regression}
\end{figure*}

Fig. \ref{fig:complete_regression} shows the full regression results for predicting next-month engagement, with model specifications in Table \ref{tab:model_details}. The results indicate that participation centralization is both statistically and practically significant in predicting next month’s engagement in all group ideologies.

\begin{table}[h]
\centering
\small
\caption{Model details across group categories.}
\label{tab:model_details}
\begin{tabular}{lcccc}
\hline
\textbf{Category} & \textbf{Model} & \(\mathbf{N^*}\) & \textbf{Pseudo-R\(^2\)} & \textbf{Dispersion} \\
\hline
Overall       & Neg. Binomial & 1760 & 0.9120 & 1.690 \\
Islamophobic  & Neg. Binomial & 435  & 0.8039 & 0.549 \\
Anti-Semitic  & Neg. Binomial & 857  & 0.9273 & 1.821 \\
Other-hate    & Neg. Binomial & 477  & 0.9726 & 1.463 \\
\hline
\end{tabular}
\\[1ex]
\parbox[t]{\textwidth}{\footnotesize \(^*\) after outlier removal.}
\end{table}

Features definitions:
\begin{itemize}
    \item \textbf{Participation Centralization}: Gini ratio for user posting activity.
    \item \textbf{Framing/Topic Homogeneity}: Gini ratio for framing/topic, calculated from the number of posts for each framing/topic.
    \item \textbf{Network Homophily}: Homophily Index.
    \item \textbf{Degree}: Number of unique hate groups affiliated with it.
    \item \textbf{Cross-group Users' Posts}: Proportion of posts created by users who posted in multiple groups.
    \item \textbf{Cross-Category Users' Posts}: Proportion of posts created by users who posted in multiple groups across different group categories/ideologies.
    \item \textbf{Framing--/Topic--}: Proportion of posts for each framing/topic.
    \item \textbf{Sentiment--}: Proportion of posts with a particular sentiment. Sentiment analysis was performed using the finiteautomata/bertweet-base-sentiment-analysis model \cite{perez2021pysentimiento}.
\end{itemize}

\subsection{Paraphrasing Narrative Examples}

To protect privacy, narrative examples in this paper were paraphrased using ChatGPT (https://chatgpt.com/) while logged into a personal account to prevent accidental exposure of the content. 
Prompt example: \textit{"Paraphrase the following hate post without changing its meaning. You may reword the adjectives, but do not remove the Islamophobia or alter the religious framing (i.e., promoting the concept of a religious duty or obligation to hate Muslims)."}

\subsection{Materials and Code Availability}
To protect user privacy and comply with Meta’s policies, raw content data cannot be shared. To support transparency and reproducibility, key aggregated statistics—including group identifiers—along with other relevant resources are available at: http://bit.ly/3KIWNoV.

\subsection{Asset Licenses}

RoBERTa large: MIT License; LLaMA 2: LLaMA 2 Community License; finiteautomata/bertweet-base-sentiment-analysis: Non-commercial, research-only license; BERTopic: MIT License.

\subsection{Extremist Narrative Framing Classification}
\subsubsection{Training Data Before and After Augmentation.}

As the initial classifier experiment resulted in low performance, we augmented the positive samples using synthetic samples Llama-2. This augmentation resulted in an improvement of the performance (see Methodology: Sample augmentation). The number of positive samples before and after augmentation for each framing strategy is available in Table \ref{tab:narrative_samples}.  The total number of annotated samples before augmentation is 940. For \textit{Us Vs. Them}, \textit{Dehumanization and Demonization}, \textit{Victimization}, and \textit{Religious}, no samples are generated for augmentation purposes. For the other four frames, the positive samples are five-times up-sampled with the generated data.

\subsection{Test Samples}
Table \ref{tab:narrative_samples} also provides the information of the number of test sample we used to evaluate the narrative framing classifiers.

\begin{table}[!htb]
\centering
\small
\begin{tabular}{@{}p{2.75cm}ccr@{}}
\toprule
\textbf{Narrative Frame} & \multicolumn{2}{c}{\textbf{\#Positive Samples}} & \textbf{\#Test} \\ 
\cmidrule(lr){2-3}
 & \textbf{Before} & \textbf{After} & \textbf{Samples} \\ \midrule
Us vs. Them & 204 & 204 & 49 \\
Dehumanization \& \\ Demonization & 239 & 239 & 339 \\
Moral Justification & 20 & 100 & 66 \\
Victimization & 104 & 104 & 30 \\
Imminent War/Crisis & 18 & 90 & 31 \\
Legitimacy & 31 & 155 & 112 \\
Heroic & 37 & 185 & 126 \\
Religious & 51 & 51 & 21 \\ \bottomrule
\end{tabular}
\caption{Number of positive samples before and after augmentation and the number of test samples for each narrative frame.}
\label{tab:narrative_samples}
\end{table}

\subsubsection{Comparison of Narrative Classifiers.}

We developed a classification pipeline combining zero-shot learning with Llama-2 and supervised fine-tuning of a transformer model, where Llama-2 served as the baseline—achieving strong recall but often overestimating frame presence, leading to lower precision. To improve accuracy, we fine-tuned a RoBERTa-large model pre-trained on online hate speech corpora. Each frame was treated as an independent binary task, assigning texts probabilities for frame membership. The dataset was split into training (70\%), validation (15\%), and test (15\%) sets. The model was trained for five epochs using cross-entropy loss and the AdamW optimizer (learning rate: $2e^{-5}$), with early stopping based on validation performance. 

Table \ref{tab:classification_singlecol} shows the comparison between baseline Llama-2 model and fine-tuned RoBERTa model in classifying hate narrative frames. The fine-tuned RoBERTa model consistently outperformed the baseline model across framings in terms of accuracy, F1 score, and precision, while the recall values were comparable. 

Misclassification of posts by the final model into incorrect narrative frames (e.g., labeling a \textit{Heroic} post as \textit{Legitimacy}) could slightly bias aggregate statistics such as frame prevalence and homogeneity within groups. However, the classification model demonstrates high performance across multiple metrics (Accuracy, F1, Precision, and Recall), and our analyses aggregate over many posts and groups. Therefore, minor classification errors are unlikely to substantially alter the observed patterns, making the analysis moderately tolerant to misclassification.


\begin{table}[h]
\centering
\small
\caption{Hate narrative classification performances of baseline Llama-2 model \cite{touvron2023llama} and fine-tuned RoBERTa model \cite{Liu2019RoBERTa}. Metrics for Llama-2 are listed on top, and RoBERTa metrics are listed below for each narrative frame.}
\label{tab:classification_singlecol}
\begin{tabular}{@{}p{1.75cm}p{6.3cm}@{}}
\toprule
\textbf{Narrative Frame} & \textbf{Metrics} \\ \midrule

Us vs. Them & 
\textbf{Llama-2}: Acc 0.50, F1 0.67, Prec 0.50, Rec 1.00 \\
& \textbf{RoBERTa}: Acc 0.87, F1 0.88, Prec 0.94, Rec 0.82 \\ \midrule

DehumDemon & 
\textbf{Llama-2}: Acc 0.52, F1 0.68, Prec 0.51, Rec 1.00 \\
& \textbf{RoBERTa}: Acc 0.88, F1 0.90, Prec 0.91, Rec 0.90 \\ \midrule

Moral Justification & 
\textbf{Llama-2}: Acc 0.49, F1 0.63, Prec 0.50, Rec 0.87 \\
& \textbf{RoBERTa}: Acc 0.88, F1 0.92, Prec 0.93, Rec 0.90 \\ \midrule

Victimization & 
\textbf{Llama-2}: Acc 0.53, F1 0.65, Prec 0.57, Rec 0.95 \\
& \textbf{RoBERTa}: Acc 0.85, F1 0.85, Prec 0.85, Rec 0.85 \\ \midrule

Imminent War/Crisis & 
\textbf{Llama-2}: Acc 0.51, F1 0.67, Prec 0.50, Rec 1.00 \\
& \textbf{RoBERTa}: Acc 0.84, F1 0.90, Prec 0.82, Rec 1.00 \\ \midrule

Legitimacy & 
\textbf{Llama-2}: Acc 0.55, F1 0.67, Prec 0.53, Rec 0.93 \\
& \textbf{RoBERTa}: Acc 0.88, F1 0.92, Prec 0.88, Rec 0.96 \\ \midrule

Heroic & 
\textbf{Llama-2}: Acc 0.55, F1 0.67, Prec 0.53, Rec 0.93 \\
& \textbf{RoBERTa}: Acc 0.89, F1 0.92, Prec 0.96, Rec 0.88 \\ \midrule

Religious & 
\textbf{Llama-2}: Acc 0.57, F1 0.64, Prec 0.55, Rec 0.76 \\
& \textbf{RoBERTa}: Acc 0.93, F1 0.94, Prec 0.89, Rec 1.00 \\

\bottomrule
\end{tabular}
\end{table}

\newpage
\FloatBarrier
\subsubsection{Codebook} \label{app:codebook}

\centering
\small
\begin{tabular}{c}
\textbf{ Extremist Narrative Framing Annotation Codebook} \\
\textbf{\textcolor{red}{**Disclaimer**}}
\end{tabular}

\begin{boxC}
This codebook is subject to an academic study that includes the presentation and analysis of extremist languages. \textbf{Readers are advised that the content may include sensitive, potentially offensive, and controversial language and ideas.} Judgment is advised while interpreting the materials, keeping in mind the academic oriented nature of this material.

The inclusion of these materials is for research and analytical purposes only. The authors and the institution do not endorse, support, or promote the extremist views and ideologies presented in these materials. The aim is to provide an academic perspective on the nature and impact of extremist languages, contributing to an understanding of these phenomena in a linguistic context. 
\end{boxC}

\raggedright

\textbf{Purpose}: The Narrative Framing Annotation Codebook stands as a guide for annotators engaged in the intricate task of categorizing extremists’ narrative frames within textual data. Designed to facilitate an understanding of how narratives are presented, this codebook offers a structured approach for identifying and annotating language use and expression. Its purpose is to enable a consistent and comprehensive analysis of narratives across various texts, ensuring that the subtleties of expression are captured and labeled.

\textbf{Scope}: Textual data from online sources, including social media and online forums. The annotation unit is one or multiple sentences (referred to as the "target") with its context in terms of preceding and subsequent sentences of the annotation target.

\textbf{Format of the codebook}: The codebook describes the inclusion criteria of a sentence to be considered under certain narrative frames. There are 8 frames commonly used by extremists that we list in this codebook (details below). For every listed narrative frame, there are 1-3 enlisted criteria, each consisting of a definition and an example. The target sentences in examples are in bold while their contexts are in plain font. The key clues that correspond to the definitions in each target sentence are highlighted in red. (In some cases, there may be only one target sentence and no context sentences.)

\textbf{Annotator Instructions}: The annotator should 1) read through this codebook and understand the inclusion criteria under each narrative frame; 2) when annotating a sample, compare it to each Narrative Frame and look for matched criteria; and 3) select all Narrative Frames categories that have matched criteria.

{ \textbf{Us Vs. Them}}

The "Us vs. Them" extremism narrative is a form of rhetoric that divides the world into two opposing groups, typically perceived as inherently good ("Us") and inherently bad ("Them"). This narrative is often used to promote extreme ideologies and violent actions. 
As examples, consider:

a. Elevate individual behaviors and characteristics to group level
\begin{boxB}
What about \textcolor{red}{all} the tens of millions of negroes that are nothing more than criminals and parasites that do nothing but breed more criminals and parasites?
\end{boxB}
b. Use of pronouns (we/they) with names assigned to groups (e.g. Jews)
\begin{boxB}
Well then , that makes the fact \textcolor{red}{these scum} are sending millions of \textcolor{red}{their kin} to \textcolor{red}{my country} , taking jobs from the White inhabitants , killing our people and mongrelizing our race all alright then .
\end{boxB}
c. Imply uncompromisable conflict/disagreement between the groups
\begin{boxB}
This whole traditionalist talking point is beyond idiotic, many people on the right do not have kids. In fact \textcolor{red}{many of the people destroying our country are fucking niggers and illegal spics with 8 fucking kids} or more. So stop and ask yourself wtf are you even talking about morons. LMAO
\end{boxB}

{ \textbf{Dehumanization and Demonization}}

The "Dehumanization and Demonization" extremism narrative is a specific form of rhetoric used to defame and marginalize a group of people, making them seem less human, inferior, or inherently evil. This narrative is often employed in the context of extremist ideologies and conflicts. 
As examples, consider:

a. Use discriminative descriptions on a group
\begin{boxB}
Here is a video of the event that is worth watching just to see a beautiful White lady screaming \textcolor{red}{`` scum , scum , scum , f * * * ing scum !!! '' at the muslim filth.}
\end{boxB}
b. Suggest a group is inferior (in intelligence, ability, moral) as a whole
\begin{boxB}
What about all the tens of millions of negroes that are \textcolor{red}{nothing more than criminals and parasites that do nothing but breed more criminals and parasites?}
\end{boxB}
c. Use analogies/sarcasms to describe a group of people is inhuman/evil
\begin{boxB}
I was beaten and robbed by two natives last year and over the last couple of years I have watched downtown turn into black , brown and yellow town. the food court at Portage place has started to \textcolor{red}{remind me of the creature cantina from star wars.}
\end{boxB}

{ \textbf{Moral Justification of Violence}}

The "Moral Justification of Violence" extremism narrative is a rhetorical framework (often inspired by extreme belief  and value systems) used by individuals or groups to justify the use of violence based on moral reasoning. This narrative is often found in extremist ideologies, whether political, religious, or social. 
As examples, consider:

a. Leverage mainstream-conflicting ideological moral to justify violence to a group or an individual from it
\begin{boxB}
\textcolor{red}{jews: 'Let's cut their funding! Filthy goyim! HAHAHA! DIE! DIE! DIE!'} *Autistic shrieking* also jews: 'NOOO! Save us! They are going to attack us!' \textcolor{red}{*Hides knives behind back*}
\end{boxB}
b. Use sophistry or propose fallacious arguments to deny, cover, or justify historical or hypothetically happened violence done to a group
\begin{boxB}
There were no extermination or death camps, there were labour camps. \textcolor{red}{Jews died for the same reasons German citizens died: typhoid, and starvation from relentless allied bombing.} You’ve see photos of emaciated bodies and immediately assume, like a good fuckimg sheep, that those we Jews killed by Hitler. \textcolor{red}{All claims of ovens and gas chambers have been debunked.} This is simple shit, dumm
\end{boxB}

{ \textbf{Victimization}}

The "Victimization" extremism narrative is a framework used by individuals or groups to portray themselves or their group as victims of crime, injustice, or oppression. This narrative is commonly employed in extremist ideologies and movements. 
As examples, consider:

a. Claim the author/speaker or the group he/she is in, is a victim of consequences caused by a group/a representative individual from a group
\begin{boxB}
\textcolor{red}{I was beaten and robbed by two natives} last year and over the last couple of years I have watched downtown turn into black , brown and yellow town. the food court at Portage place has started to remind me of the creature cantina from star wars .
\end{boxB}
b. Claim there are real or hypothetical negative social consequences inflicted by a group
\begin{boxB}
Dems have shown themselves to be deranged, delusional, insane people who want to destroy our America!  Will we allow this?  \textcolor{red}{They shoot our people with snipers, then kill our Americans at concerts, they would gladly kill us all to take what we have.}  God bless you and God bless the USA...
\end{boxB}

{ \textbf{Imminent War/Crisis}}

The "Imminent War/Crisis" extremism narrative is a rhetorical strategy used by individuals or groups, often with extremist ideologies, to convey a sense of impending large-scale conflict if their ideology is not adopted by the world. This narrative is designed to inflict urgency, fear, and a sense of inevitability about an incoming crisis.
As examples, consider:

a. Suggest conflicts between two groups are uncompromisable, ignoring which will lead to an intense danger
\begin{boxB}
\textcolor{red}{``If France won't address the issues of Islam — I think that country could just fall into civil war,''}   Ayaan Hirsi Ali: Tackle Islam or face civil war
\end{boxB}
b. Suggest a (physical or ideological) war between groups is the only solution to solve conflicts or the crisis
\begin{boxB}
Let's give Muslims the \textcolor{red}{choice to convert to a real religion of peace, and if they do not, deport them to a Muslim country. This is a war and the West is getting its butt kicked by these savages and their traitorous left wing allies, and the \#NWO}
\end{boxB}
c. Suggest a action to be taken against a group, in response to an imminent threat from that group
\begin{boxB}
\#Cucked PROGS want to enable this anarchy here. \textcolor{red}{Time to \#Purge this mindset and those who enable it.}
\end{boxB}

{ \textbf{Legitimacy}}

The "Legitimacy" extremism narrative is a narrative frame used by individuals or groups with extremist views, to establish and justify the validity and righteousness of their actions, or ideology. This narrative seeks to describe their stance as legitimate, reasonable.
As examples, consider:

a. Align group (that the author/speaker leaned on) behaviors or ideology to (outdated or mainstream-contradicting) figures, texts, or symbols
\begin{boxB}
God over man, man over woman, women over children. This is the order of god. Break it at your peril and down fall. \textcolor{red}{Play by mother nature’s rules or she will come give you a spanking. Without good women in the home, nurturing, teaching, loving and inspiring our children we are doomed. “It takes two to make a thing go right”. With out the man at the head of a household, protecting, defending and supporting the family we are doomed. For evil to flourish, good men must do nothing.}
\end{boxB}

{ \textbf{Heroic}}

The "Heroic" extremism narrative is a form of rhetoric used by people with extremist ideologies, to frame their actions, struggles, or causes as heroic, and deserving of admiration. This is designed to romanticize and glorify the group or its actions, making them appealing, especially to potential recruits or sympathizers.
As examples, consider:

a. Romanticize group (that the author/speaker leaned on) actions or beliefs, portraying them as heroic responses to a situation or context
\begin{boxB}
Our cause is \textcolor{red}{not just a cause}; it is a \textcolor{red}{crusade against the tide of non-white complacency and mediocrity that has gripped our society.}
\end{boxB}
b. Promote a charismatic leader who embodies the ideals of the movement and serves as a role model
\begin{boxB}
\textcolor{red}{Our leader is not just a person, but a symbol of our collective resolve} of the muslim TROUBLES. \textcolor{red}{A leader who embodies the courage, the vision, and the unyielding strength that our cause demands. A leader who stands as a beacon of hope in these troubled times.}
\end{boxB}
c. Emphasize personal sacrifice for a group ideology, including sacrificing personal effort, safety, social standing, and life.
\begin{boxB}
Your hate and selective persecution makes YOU the enemy. \textcolor{red}{Men are the heroes, and chivalric knights who protect you. that's why you hate them.} we see you for the enemy scum.
\end{boxB}

{ \textbf{Religious}}

The "Religious Extremism" narrative is a type of narrative frame used by certain groups or individuals who interpret religious texts, and beliefs to support, justify, or promote extremist ideologies and actions. This narrative is often characterized by its intolerance, and often hostility towards those outside of or opposed to their religious viewpoint.
As examples, consider:

a. Use a particular interpretation of religious texts or beliefs to exclusively claim truth statements
\begin{boxB}
\textcolor{red}{God over man, man over woman, women over children. This is the order of god. Break it at your peril and down fall.} Play by mother nature’s rules or she will come give you a spanking. Without good women in the home, nurturing, teaching, loving and inspiring our children we are doomed. “It takes two to make a thing go right”. With out the man at the head of a household, protecting, defending and supporting the family we are doomed. For evil to flourish, good men must do nothing.
\end{boxB}

\end{document}